\documentclass[12pt]{article}
\usepackage{epsfig}
\usepackage{graphicx}
\usepackage{a4}
\usepackage{amsmath}
\usepackage{latexsym}
\usepackage{cite}
\usepackage{lineno}
\usepackage{xspace}

\usepackage{color}
\usepackage{colordvi}

\graphicspath{{Pics/}}
\DeclareGraphicsExtensions{.eps,.ps}

\usepackage{pslatex}
\usepackage[latin1]{inputenc}
\usepackage[T1]{fontenc}
\usepackage{amssymb}
\usepackage{url}

\usepackage{geometry}
\newcommand{\msbar}{$\overline{\text{MS}}\, $\xspace}

\begin{document}

\begin{titlepage}
\noindent
\\
DESY 19-211 
\\

\vspace{1.0cm}

\begin{center}
  {\bf 

\large

Improved constraints on parton distributions using LHCb, ALICE and HERA heavy-flavour measurements and implications for the predictions for prompt atmospheric-neutrino fluxes 
  }
  \vspace{1.0cm}

  {\large
    PROSA Collaboration
  }\\

  \vspace{0.4cm}

\end{center}
\noindent
O.~Zenaiev$^{1a*}$, \mbox{M.V.~Garzelli}$^{2}$, K.~Lipka$^{3}$, \mbox{S.-O.~Moch}$^{1}$, A.~Cooper-Sarkar$^{4}$, F.~Olness$^{5}$, A.~Geiser$^{3}$, G.~Sigl$^{1}$\\

\noindent

{\footnotesize{
		\noindent
		$^{1}$II. Institut f\"ur Theoretische Physik, Universit\"at Hamburg, Luruper Chaussee 149, D-22761 Hamburg, Germany \\
		$^{2}$Dipartimento di Fisica e Astronomia, Universit\'a degli Studi di Firenze, \& INFN, Firenze, Italy, \& Institut f\"ur Theoretische Physik, Eberhard Karls Universit\"at T\"ubingen, T\"ubingen, Germany \\
		$^{3}$DESY, Notkestrasse 85, D-22607 Hamburg, Germany\\
		$^{4}$University of Oxford, Keble Road, Oxford OX1 3RH, U.K.\\
		$^{5}$Southern Methodist University, Box 0175 Dallas, TX 75275-0175, U.S.A.\\
		$^{a}$now at CERN, CH-1211 Geneva 23, Switzerland\\
    
    \noindent
		$^{*}$oleksandr.zenaiev@cern.ch\\
	}
}

  \vspace{1.0cm}
\begin{center}
\large
{\bf Abstract}
\vspace{-0.2cm}
\end{center}
The impact of measurements of heavy-flavour production in deep inelastic $ep$ scattering and
in $pp$ collisions on parton distribution functions is studied in a QCD analysis at next-to-leading order. 
Recent combined results of inclusive and heavy-flavour production cross sections in deep inelastic scattering at HERA are 
investigated together with heavy-flavour production measurements at the LHC. Differential cross sections of charm- and 
beauty-hadron production measured by the LHCb collaboration at the centre-of-mass energies of 5, 7 and 13 TeV as well 
as the recent measurements of the ALICE experiment at the centre-of-mass energies of 5 and 7 TeV are explored. 
These data impose additional constraints on the gluon and the sea-quark distributions at low partonic fractions
$x$ of the proton momentum, down to $x\approx10^{-6}$. The impact of the resulting parton distribution function in the predictions for the prompt atmospheric-neutrino fluxes is studied.

\vfill
\end{titlepage}

%
%
\newpage

\section{Introduction}
\label{sect:intro}

The fundamental structure of the nucleon is described by the theory of strong interactions, quantum chromodynamics (QCD).
In the collinear factorisation, the nucleon structure is expressed in terms of parton distribution functions (PDFs), defined 
as probability densities for partons to carry a fraction $x$ of the nucleon momentum at a factorisation scale $\mu_f$. While the 
scale evolution of the PDFs is calculated in perturbative QCD (pQCD) using the DGLAP equations~\cite{Dokshitzer:1977sg,Gribov:1972ri,Altarelli:1977zs,Curci:1980uw,Furmanski:1980cm,Moch:2004pa,Vogt:2004mw}, the 
$x$ dependence must be constrained from the experimental measurements. The constraining power of experimental data 
on particular parton distributions is to a large extent defined by the acceptance of the experiment. Measurements of 
neutral current (NC) and charged current (CC) cross sections in deep inelastic scattering (DIS) at HERA~\cite{Abramowicz:2015mha} probe the $x$ range of $10^{-4}<x<10^{-1}$, impose most significant constraints on the light quark PDFs and probe the gluon distribution via scaling violations. Additional constraints on the flavour separation of the quark sea and on the gluon distribution at low and high $x$ are obtained by using the measurements from fixed target experiments and in proton-(anti)proton collisions. 
Heavy-flavour production in proton-proton ($pp$) collisions at the LHC is dominated by gluon-gluon fusion, therefore corresponding measurements probe the gluon distribution directly~\cite{Zenaiev:2015rfa,Gauld:2015yia,Gauld:2016kpd,Bertone:2018dse}. 
The measurements of forward charm~\cite{Aaij:2013mga} and beauty~\cite{Aaij:2013noa} production by the LHCb experiment at the centre-of-mass energy $\sqrt{s}=7$ TeV were used for the first time by the PROSA collaboration~\cite{Zenaiev:2015rfa} to improve constraints on the gluon distribution at $5 \times 10^{-6}< x < 10^{-4}$, in the region hardly covered by any other measurements to that date. 
The resulting PDFs (PROSA 2015) were further used to predict the prompt neutrino flux from the decays of charmed mesons produced via cosmic ray interactions in the Earth's atmosphere~\cite{Garzelli:2016xmx}, which constitute an irreducible background in searches for the extraterrestrial neutrino flux by IceCube. 
{The first QCD analysis~\cite{Zenaiev:2015rfa} using forward heavy-flavour production at the LHCb has triggered attention of several PDF groups, which confirmed the significant improvement of the constraints on the gluon distribution by using forward heavy-flavour production measurements obtained with the LHCb experiment. In particular, one group has updated the PDF set NNPDF3.0~\cite{Ball:2014uwa} by including the LHCb charm measurements into the NNPDF3.0 fit using the Bayesian reweighting method and providing the NNPDF3.0+LHCb PDF set~\cite{Gauld:2015yia}. Further LHCb measurements collected at $\sqrt{s}=5$ TeV, 7 TeV and 13 TeV were included in the PDF studies of Refs.~\cite{Gauld:2016kpd,Bertone:2018dse}, where the normalised distributions and cross-section ratios between measurements at different centre-of-mass energies were used, following the suggestion of Ref.\cite{Cacciari:2015fta}, to quantify the impact of the LHCb $D$-meson measurements on the NNPDF3.0 and NNPDF3.1 PDF sets. In all analyses~\cite{Gauld:2015yia,Gauld:2016kpd,Bertone:2018dse}, the pQCD predictions for charm production in the forward region based on the FONLL approach~\cite{Cacciari:1993mq} were used.}

Recent improvements in the precision of the HERA measurements~\cite{Abramowicz:2015mha,H1:2018flt}, new experimental data on heavy flavour production at the LHC at 
different $\sqrt{s}$~\cite{Aaij:2015bpa,Aaij:2016jht,Acharya:2017jgo,Acharya:2019mgn}, together with new developments in the theory and improvements of the phenomenological tools, offer 
possibilities for stronger constraints on the gluon distribution at low $x$. These improvements in experimental measurements 
and the theory are explored in the QCD analysis presented in this paper, which updates the earlier PDF result~\cite{Zenaiev:2015rfa}. The results, referred to as PROSA 2019, are used to update the predictions for the prompt atmospheric-neutrino fluxes.

\section{Input data sets and used theory predictions}
\label{sec:qcdanalysis}

The main objective of the present QCD analysis is to demonstrate the constraining power of the updated measurements of 
heavy-flavour production in DIS and $pp$ collisions for the determination of the PDFs of the proton. 
The QCD analysis is performed at next-to-leading order (NLO) using the xFitter framework~\cite{Alekhin:2014irh}. 
The updated combinations of the inclusive DIS cross sections~\cite{Abramowicz:2015mha} and of charm and beauty production cross sections~\cite{H1:2018flt} are used together with the measurements of charm and beauty hadroproduction in $pp$ scattering at the LHC. The latter include the measurements of charm hadroproduction by the LHCb collaboration at $\sqrt{s}$ = 5 TeV~\cite{Aaij:2016jht}, 7 TeV~\cite{Aaij:2013mga} and 13 TeV~\cite{Aaij:2015bpa}, and by ALICE at $\sqrt{s}$ = 5 TeV~\cite{Acharya:2019mgn} and 7 TeV~\cite{Acharya:2017jgo}. The measurements of beauty hadroproduction by LHCb at $\sqrt{s}$ = 7 TeV~\cite{Aaij:2013noa} are also used.

\begin{figure}
    \centering
    \includegraphics[width=1.00\textwidth]{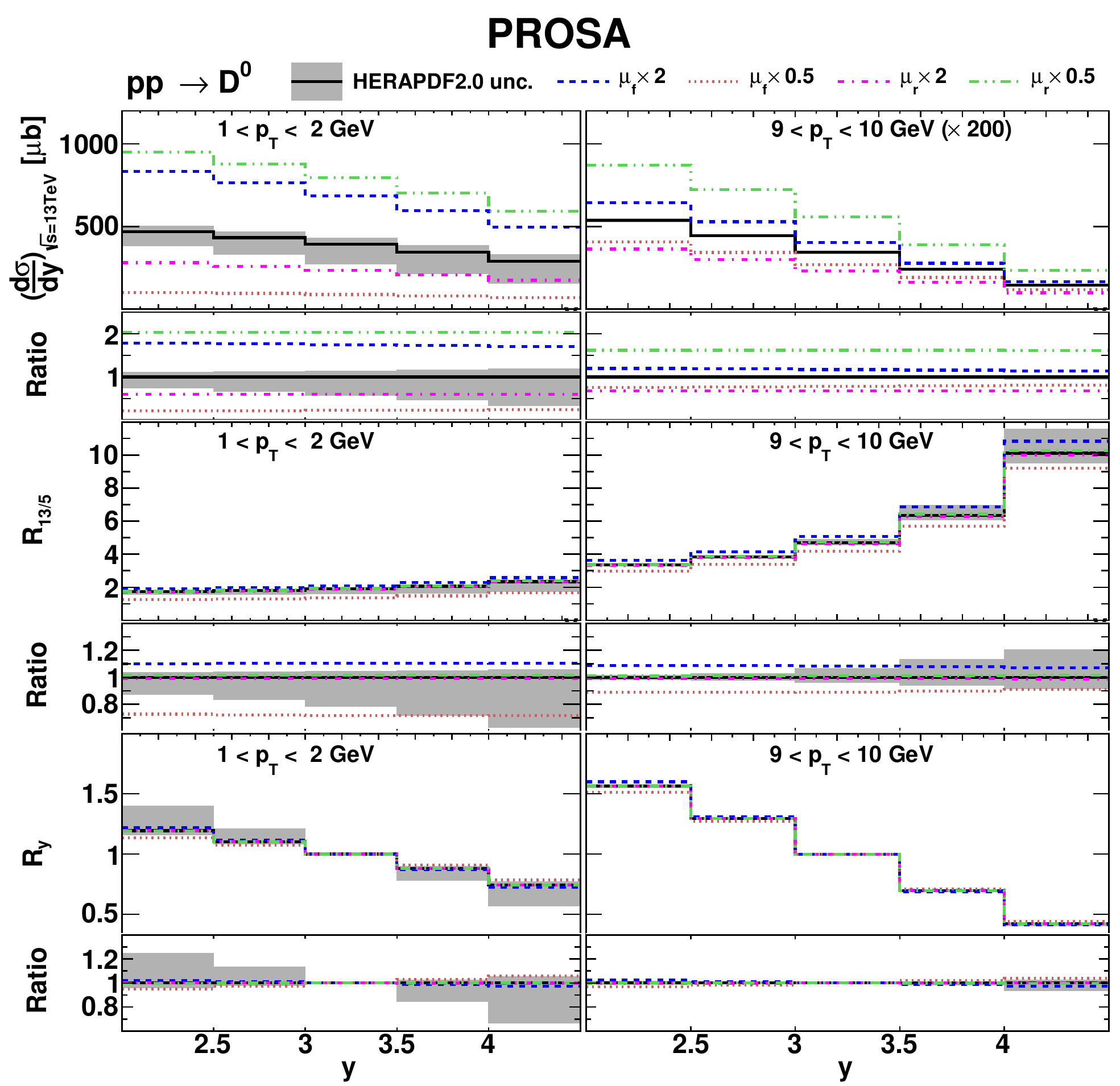}
    \caption{NLO QCD predictions for $D^{0}$ meson production at LHCb at $\sqrt{s} = 13$~TeV (top row), 
      cross section ratios at $\sqrt{s} = 13$~TeV and $\sqrt{s} = 5$~TeV, denoted as $R_{13/5}$ (middle row), and normalised cross sections, denoted as $R_y$ (bottom row) 
      with their PDF uncertainties obtained using the HERAPDF2.0 PDF set with three active flavours (\texttt{HERAPDF20\_NLO\_FF3A\_EIG} and \texttt{HERAPDF20\_NLO\_FF3A\_VAR} LHAPDF sets) and different scale choices for two representative $p_T$ regions 
      (the cross sections in the top right panel are scaled by a factor of $200$ for presentation purposes).
      The lower panels indicate the ratio of the predictions to the nominal one.}
    \label{fig:scalevars}
\end{figure}

The cross sections measured by LHCb and ALICE in each $p_T$ range are normalised in rapidity $y$, $\frac{{\rm d}^2\sigma}{{\rm d}y{\rm d}p_T} / \left(\frac{{\rm d}^2\sigma}{{\rm d}y{\rm d}p_T}\right)_{0}$. Here, $\left(\frac{{\rm d}^2\sigma}{{\rm d}y{\rm d}p_T}\right)_{0}$ is the cross section in the central LHCb rapidity bin, $3 < y < 3.5$. 
{This bin is chosen for normalisation since all relevant LHCb data sets have a measurement in this rapidity range.
Choosing a different bin for normalisation would not change the results.}
When normalising cross sections in this way, ALICE measurements at $|y| < 0.5$ are divided by the LHCb cross-section measurement in $3 < y < 3.5$.
The advantage of using the normalised cross section, demonstrated in the earlier PROSA analysis~\cite{Zenaiev:2015rfa}, 
is a significant reduction of the scale dependence in the theoretical prediction,  while retaining the sensitivity to 
the PDFs{, as illustrated in Fig.~\ref{fig:scalevars}. 
As shown in the same figure, this observable exhibits a smaller scale dependence than ratios of cross sections at different centre-of-mass energies which were used in Refs.~\cite{Gauld:2016kpd,Bertone:2018dse}.}

In the presented QCD analysis, bin-to-bin correlations in the input measurements are taken into account as described in the following. The treatment of correlated experimental uncertainties for the HERA data follows that of the original publications~\cite{Abramowicz:2015mha,H1:2018flt}.
 
The correlated uncertainties in the ALICE and LHCb measurements reported in the original publications~\cite{Aaij:2016jht, Aaij:2013mga, Aaij:2015bpa, Acharya:2019mgn, Acharya:2017jgo, Aaij:2013noa} and listed in the respective tables as exact uncertainty values in each kinematic bin in $p_T$ and $y$, 
are treated as fully correlated, and the uncorrelated uncertainties are obtained by subtracting the correlated ones from the 
total uncertainties, in quadrature. Because of this treatment of systematics, most of these correlated systematic uncertainties cancel in the calculation of the normalised cross sections. In case of the LHCb cross section ratio measurements, the uncertainties cancel completely. Further systematic uncertainties, reported as error intervals, see e.g. Table~(2) of Ref.~\cite{Aaij:2016jht}, are assumed uncorrelated, since no details about their size in individual $p_T$ and $y$ bins are provided. 
For different final state measurements within one experiment, the tracking and luminosity uncertainties are treated as correlated. 
Furthermore, all experimental uncertainties are treated as uncorrelated among measurements at different centre-of-mass energies. 
The uncorrelated uncertainties in the normalised cross sections $\left(\frac{{\rm d}^2\sigma}{{\rm d}y{\rm d}p_T}\right)_{0}$ are propagated as correlated uncertainties to the respective complementary rapidity bins.
It is worthwhile to note, that the details of the experimental uncertainties and their correlations in each individual kinematic range is of great importance and therefore most detailed information about the systematic correlations in the experimental measurement is required.

In the presented QCD analysis, the scale evolution of partons is calculated through the DGLAP equations at NLO, as implemented in 
the QCDNUM programme~\cite{Botje:2010ay}.  
The description of the inclusive HERA data in the PDF fits improves in the kinematic range of small $x$ and low virtuality $Q^2$, by including higher twist effects~\cite{Accardi:2016ndt,Alekhin:2017kpj}
or, alternatively, small $x$ resummation~\cite{Ball:2017otu,Abdolmaleki:2018jln}. 
These upgrades are left for future analyses, once all necessary theoretical ingredients have become available. 
The changes in the PDFs when varying the $Q^2_{\text{min}} = 3.5~\textrm{GeV}^2$ cut imposed on the HERA data, $2.5 \leq Q^2_\textrm{min}\leq 5.0~\textrm{GeV}^2$, are found to be small with respect to other uncertainties. 
Therefore we are confident that inclusion of higher-twist terms would not modify the results of this analysis in a substantial way.

The theoretical predictions for the heavy-quark and inclusive HERA data are obtained using the OPENQCDRAD~\cite{openqcdrad} code in the 
fixed-flavour-number scheme (FFNS) with three active flavours in the proton using the \msbar mass scheme, following Ref.~\cite{H1:2018flt}. 
Similar to the earlier PROSA analysis~\cite{Zenaiev:2015rfa}, the theoretical predictions for the fully differential 
heavy-quark hadroproduction in $pp$ collisions, available at NLO in FFNS, are used. These are calculated using 
the MNR code~\cite{Mangano:1991jk}, with the single-particle inclusive distributions computed using the pole mass scheme for the heavy quarks, 
and translated into the \msbar mass scheme expressions using the \msbar mass $m_Q(m_Q)$ and following Ref.~\cite{Dowling:2013baa}.
The \msbar mass scheme is then consistently used in the calculations for all used processes.

The factorisation and renormalisation scales are chosen to be $Q$ for inclusive DIS, and $\mu_r = \mu_f = \sqrt{Q^2 + 4m_Q(m_Q)^2}$ for heavy quark production in DIS, respectively, with $m_Q(m_Q)$ representing the heavy-quark mass in the \msbar scheme. 
For heavy quark production in $pp$ collisions, $\mu_r = \mu_f = \sqrt{4m_Q(m_Q)^2+p_T^2}$ is assumed. 

The calculations for heavy quark hadroproduction are supplemented with phenomenological non-perturbative fragmentation 
functions to describe the transition of heavy quarks into hadrons. The fragmentation of charm quarks into $D$ mesons is 
described by the Kartvelishvili function $D_Q(z) \propto z^{\alpha_K}(1-z)$ 
with $\alpha_K = 4.4 \pm 1.7$ as measured at HERA~\cite{Aaron:2008ac,Chekanov:2008ur}, 
and for the fragmentation of beauty quarks $\alpha_K = 11 \pm 3$ is used as measured at LEP~\cite{Nason:1999zj}, following 
the previous PROSA analysis~\cite{Zenaiev:2015rfa}.
Studies of the uncertainties related to the fragmentation in Ref.~\cite{Alekhin:2012un} 
for a determination of the charm-quark mass in the \msbar scheme from
deep inelastic scattering at HERA data have shown that the dominant effect is
captured by varying $\alpha_K$ within its uncertainties. 
This treatment of charm quark fragmentation is independent of the choice of a particular renormalisation scheme for the heavy quark mass. 
The latter is needed in a determination of the initial condition for the perturbative heavy quark fragmentation
function, which is known to NNLO~\cite{Melnikov:2004bm}.
The subsequent range of evolution in the case of charm quark fragmentation
into $D$ mesons from the scale of hadronisation to scales of order of the
charm quark mass is very short, so that the modelling with the non-perturbative Kartvelishvili function $D_Q(z)$ is justified.

The main QCD analysis is performed in the FFNS and the sensitivity of the heavy quark measurements to the PDFs and to the masses 
of the charm and beauty quarks is fully explored by treating $m_c(m_c)$ and $m_b(m_b)$ as free parameters in the fit.
The fit is also performed in the variable flavour number scheme (VFNS) to allow for incorporation in shower Monte Carlo event generators and applications in e.g. 
underlying event tuning at the LHC.

\section{PDF parametrisation}
\label{sec:pdfparam}

The PDFs are parametrised at the starting evolution scale of $\mu^2_{f0} = 1.9$~GeV$^2$, similar as in Ref.~\cite{Abramowicz:2015mha} and Ref.~\cite{Bonvini:2019wxf}, as follows:
\begin{equation}\begin{aligned}
xg(x) &= A_{g} x^{B_{g}}\,(1-x)^{C_{g}}\, (1 + F_{g} {\log x}),\\
xu_\mathrm{v}(x) &= A_{u_\mathrm{v}}x^{B_{u_\mathrm{v}}}\,(1-x)^{C_{u_\mathrm{v}}}\,(1+E_{u_\mathrm{v}}x^2) ,\\
xd_\mathrm{v}(x) &= A_{d_\mathrm{v}}x^{B_{d_\mathrm{v}}}\,(1-x)^{C_{d_\mathrm{v}}},\\
x\overline{\mathrm{U}}(x)&= A_{\overline{\mathrm{U}}}x^{B_{\overline{\mathrm{U}}}}\, (1-x)^{C_{\overline{\mathrm{U}}}}\, (1+D_{\overline{\mathrm{U}}}x), \\
x\overline{\mathrm{D}}(x)&= A_{\overline{\mathrm{D}}}x^{B_{\overline{\mathrm{D}}}}\, (1-x)^{C_{\overline{\mathrm{D}}}}.
\end{aligned}
\label{eq:dv}
\end{equation}

Here, $xg(x)$, $xu_{\mathrm{v}}(x)$ and $xd_{\mathrm{v}}(x)$ represent the gluon, up and down valence quark distributions, respectively. The sea quark distribution is defined as $x\Sigma(x)=x\overline{u}(x)+x\overline{d}(x)+x\overline{s}(x)$, with $x\overline{u}(x)$, $x\overline{d}(x)$, and $x\overline{s}(x)$ denoting the up, down, and strange antiquark distributions, respectively.
For the up- and down-type antiquark distributions, $x\overline{\mathrm{U}}(x)$ and $x\overline{\mathrm{D}}(x)$, relations $x\overline{\mathrm{U}}(x) = x\overline{u}(x)$ and $x\overline{\mathrm{D}}(x) = x\overline{d}(x) + x\overline{s}(x)$  are assumed.
The normalisation parameters $A_{u_{\mathrm{v}}}$, $A_{d_\mathrm{v}}$, and $A_{g}$ are determined by the QCD sum rules.
The strangeness fraction $f_{s} = x\overline{s}/( x\overline{d} + x\overline{s})$ is fixed to
$f_{s}=0.4$ as in the HERAPDF2.0 analysis~\cite{Abramowicz:2015mha}.
Additional constraints $B_{\overline{\mathrm{U}}} = B_{\overline{\mathrm{D}}}$ and $A_{\overline{\mathrm{U}}} = A_{\overline{\mathrm{D}}}(1 - f_{s})$ are imposed to ensure the same normalisation for the $x\overline{u}$ and $x\overline{d}$ distributions as $x \to 0$.
The term $F_g\log x$ was proposed in~\cite{Bonvini:2019wxf} to provide a flexible functional form at low $x$ and replace the 3-parameter extra term in Ref.~\cite{Abramowicz:2015mha}.

The predicted and measured cross sections together with their corresponding uncertainties are used to build a global $\chi^2$, minimised to determine the initial PDF
parameters. The $\chi^2$ definition follows that of Eq.~(32) in Ref.~\cite{Abramowicz:2015mha}. In the minimisation, performed 
using the MINUIT package~\cite{James:1975dr}, the experimental uncertainties in the heavy-quark normalised cross 
sections are treated as additive, and the treatment of the experimental uncertainties for the HERA DIS data follows 
the prescription given in Ref.~\cite{Abramowicz:2015mha}.

The parameters in Eq.~(\ref{eq:dv}) are selected by first parametrising each PDF as
\begin{equation}
\begin{split}
xf(x) &= Ax^B(1-x)^C(1+Dx+Ex^2+F \cdot {\log x}), ~~~f=g\\
xf(x) &= Ax^B(1-x)^C(1+Dx+Ex^2), 
~~~f=u_\mathrm{v},d_\mathrm{v},\overline{\mathrm{U}},\overline{\mathrm{D}}
\end{split}
\label{eq:de}
\end{equation}
and setting all $D$ and $E$ parameters to zero.
Additional parameters in each resulting PDF are included in the fit one at a time. 
The improvement in $\chi^2$ of the fits is monitored and the procedure is stopped when no further improvement is observed. The inclusion of the $F_{g}$ parameter does not lead to significant change in $\chi^2$, in particular, its fitted value is consistent with $0$ within uncertainty, however the variation of $F_{g}$ significantly affects the fit uncertainties.

To ensure that the gluon PDF at low $x$ is not over-constrained in the fit, different functional forms in the parametrisation were tested, as used in the ABMP16~\cite{Alekhin:2017kpj}, CT14~\cite{Dulat:2015mca}, HERAPDF2.0~\cite{Abramowicz:2015mha} and Bonvini-Giuli (BG)~\cite{Bonvini:2019wxf} PDF fits:%
\footnote{Note that in this analysis, it was not possible to achieve convergence of the fit using the MMHT2014 parametrisation~\cite{Harland-Lang:2014zoa}, because the data sets used did not have sufficient sensitivity to the gluon distribution at high $x$.}

\begin{equation}
\begin{aligned}
\textrm{ABMP16:}~~~~~~ &xg(x)=A (1 - x)^b x^{a (1 + \gamma_{1} x)},\\
\textrm{CT14:}~~~~~~ &xg(x) = Ax^{a_1}(1-x)^{a_2}(e_0(1-y)^2+e_1(2y(1-y))+y^2), \\
 ~~~~~~ &y=2\sqrt{x}-x,\\
\textrm{HERAPDF2.0:}~~~~~~ &xg(x)=A_gx^{B_g}(1-x)^{C_g}+A^{\prime}_gx^{B^{\prime}_g}(1-x)^{25},\\
\textrm{HERAPDF2.0 no flex. $g$:}~~~~~~ &xg(x)=A_gx^{B_g}(1-x)^{C_g},\\
\textrm{BG:}~~~~~~ &xg(x)=A_{g} x^{B_{g}}\,(1-x)^{C_{g}}\, (1 + F_{g} {\log x} + G_{g} {\log^2 x}),\\
\end{aligned}
\label{eq:gluonpar}
\end{equation}

\begin{figure}
    \centering
    \includegraphics[width=0.49\textwidth]{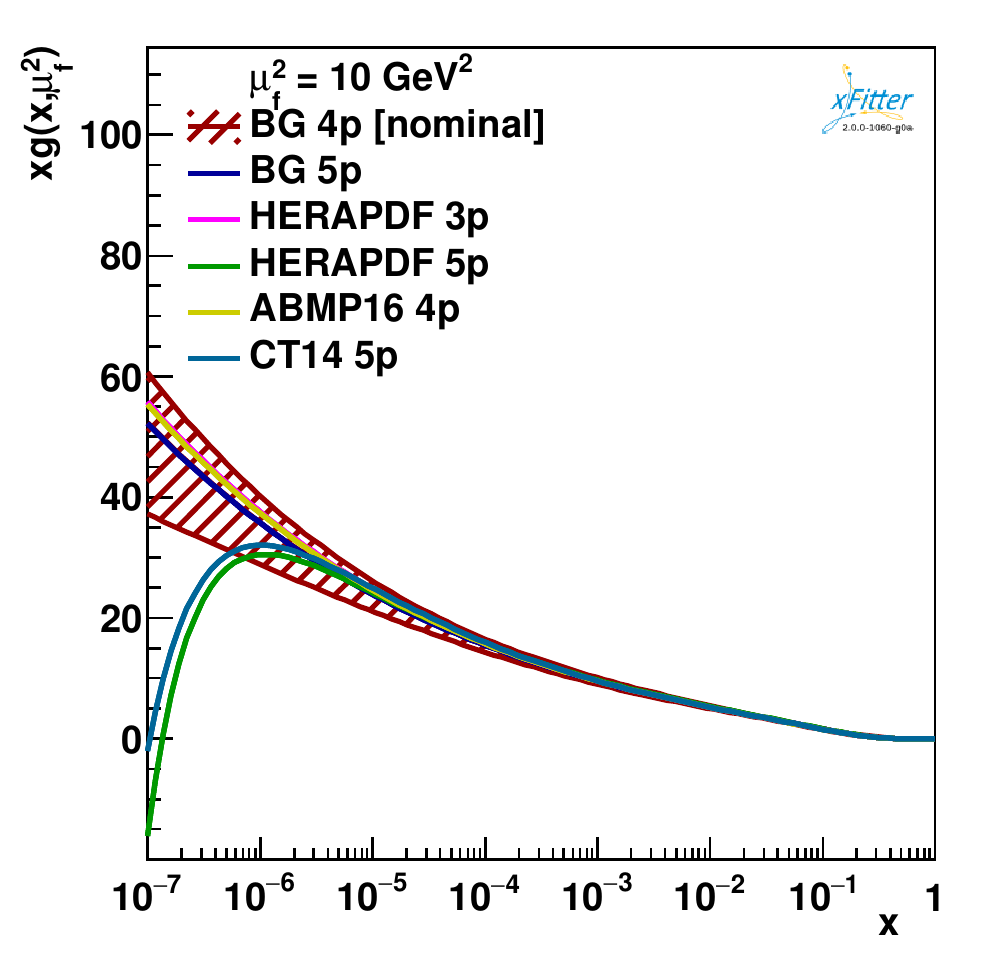}
    \includegraphics[width=0.49\textwidth]{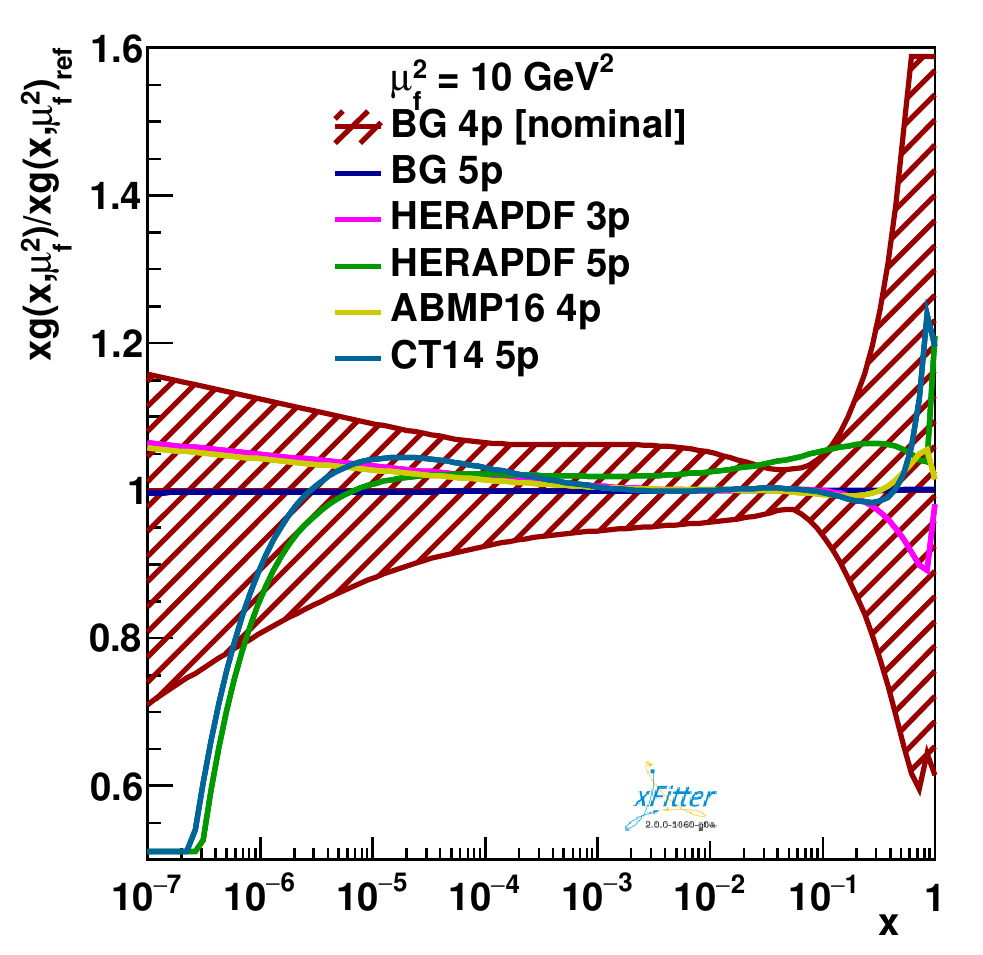}
    \caption{Left panel: the gluon PDF with their total uncertainties at the scale $\mu^2_f=10$ GeV$^2$ obtained using different gluon parametrisations, see Eq.~(\ref{eq:gluonpar}). Right panel: the same PDFs normalised to the distribution obtained using the nominal parametrisation.}
    \label{fig:gluonpar}
\end{figure}

These functional forms are characterised by $3$ (HERAPDF2.0 no 'flexible' $g$), $4$ (ABMP16) or $5$ (CT14, HERAPDF2.0, BG) parameters controlling the gluon PDF, c.f.\ $4$ parameters in the presented nominal parametrisation of Eq.~(\ref{eq:dv}). 
The resulting gluon distributions are presented in Fig.~\ref{fig:gluonpar}. The parametrisations of ABMP16, HERAPDF2.0 without the flexible gluon, and BG provide very similar results to that of the nominal parametrisation in Eq.~(\ref{eq:dv}). 
Note that also the HERAPDF2.0 analysis considered the parametrisation without the flexible gluon, referred to as an `alternative' gluon parametrisation~\cite{Abramowicz:2015mha}, provided primarily for predictions of cross sections at very low $x$, such as very high-energy neutrino cross sections.

The fit using the HERAPDF2.0 and CT14 parametrisations yielded a gluon distribution with a sharp turnover to negative values at $x \sim 10^{-6}$, i.e.\ at the edge of the kinematic reach of the used measurements. Using such PDFs would lead to a negative prediction for the total charm hadroproduction cross sections at $\sqrt{s} \gtrsim 20$~TeV, similar to the observation of Ref.~\cite{Accardi:2016ndt}. Therefore these parametrisations are discarded (despite they provide an improved $\chi^2$, by $22$ and $7$ units when using the HERAPDF2.0 and CT14 parametrisations, respectively).

\section{PDF uncertainties}
\label{sec:pdfunc}

The PDF uncertainties are investigated according to the general approach of the HERAPDF2.0 analysis~\cite{Abramowicz:2015mha}, with the fit, model, and parametrisation uncertainties taken into account.

The fit uncertainties arising from the uncertainties in the measurements are estimated by using the Hessian method, adopting the tolerance criterion of $\Delta \chi^2$ = 1, and correspond to 68\% confidence level.

To investigate the impact of model assumptions on the resulting PDFs, alternative fits are performed and the differences to the central result are considered as model uncertainties. The strangeness fraction is varied as $0.3 \leq f_{s} \leq 0.5$ and the value of $Q^2_{\text{min}}$ imposed on the HERA data as $2.5 \leq Q^2_\textrm{min}\leq 5.0~\textrm{GeV}^2$. The FFNS strong coupling constant is assumed as $0.105 < \alpha_s^{n_f=3}(M_Z) < 0.107$ (corresponding to the VFNS values of $0.117 < \alpha_s^{n_f=5}(M_Z) < 0.119$~\cite{Tanabashi:2018oca}). The variation of the fragmentation parameters $\alpha_K = 4.4 \pm 1.7$ for charm hadrons~\cite{Aaron:2008ac,Chekanov:2008ur} and $\alpha_K = 11 \pm 3$ for beauty hadrons~\cite{Nason:1999zj} is performed.
The scales $\mu_f$ and $\mu_r$ for heavy quark production are varied independently and simultaneously up and down by a factor of two, excluding variations of the two scales in opposite directions. Note that for 
the normalised cross section predictions, the simultaneous variation of the $\mu_f$ and $\mu_r$ scales in the same direction results in the largest deviation in the 
resulting PDFs and is considered as one PDF uncertainty eigenvector.

The parametrisation uncertainty is estimated by extending the functional form of each PDF in Eq.~(\ref{eq:dv}) with additional parameters $D$ and $E$, see Eq.~(\ref{eq:de}), 
which are added or removed one at a time and do not impact the $\chi^2$. 
Furthermore, the shape of the gluon PDF is extended by adding a $+G_g\log^2 x$ term~\cite{Bonvini:2019wxf}. This modification does not result in an improvement in $\chi^2$ and therefore is not considered in the nominal parametrisation. 
The variation of the starting scale, $1.6 < \mu_\mathrm{f0}^2 < 2.2$~GeV$^2$, is also taken into account as contribution to the parametrisation uncertainty. The parametrisation uncertainty is constructed at any given scale as an envelope built from the maximal differences between the PDFs resulting from all the parametrisation variations and the central fit at each $x$ value.

The total PDF uncertainty is obtained by adding experimental, model, and parametrisation  uncertainties in quadrature.


\section{PROSA 2019 parton distributions}
\label{sec:results}

The quality of the overall fit can be judged based on the global $\chi^2$ divided by the number of degrees of freedom, $n_{dof}$. For each data set included in the fit, a partial $\chi^2$
divided by the number of measurements (data points), $n_{dp}$, is provided. The correlated part of $\chi^2$ quantifies the influence of the correlated systematic uncertainties in the fit. The global and partial $\chi^2$ values for each data set are listed in Table~\ref{tab:chi}, {illustrating a satisfactory agreement among all the data sets (note that the $\chi^2$ expression does not include theoretical uncertainties, but we account for them in the evaluation of the theory uncertainties associated to the PDFs)}. The central values and the uncertainties of the fitted PDF parameters are given in Table~\ref{tab:pars}. 
The fitted masses of the heavy quarks are $m_c(m_c) = 1.230 \pm 0.031$ GeV, $m_b(m_b) = 3.977 \pm 0.100$~GeV. These values are a bit lower than, but consistent with, those obtained from the HERA data only~\cite{H1:2018flt}.
The corresponding full set of other potential systematic uncertainties was not evaluated here.

\begin{table}
\renewcommand*{\arraystretch}{1.12}
\tabcolsep1cm
    \centering
\begin{tabular}{ll}
    Data set & $\chi^2/n_{dp}$ \\
    \hline
    HERA CC $e^{+}p$ & 62 / 39  \\ 
    HERA CC $e^{-}p$ & 49 / 42  \\ 
    HERA NC $e^{-}p$ & 227 / 159  \\ 
    HERA NC $e^{+}p$ 820 GeV & 68 / 70  \\ 
    HERA NC $e^{+}p$ 920 GeV & 440 / 377  \\ 
    HERA NC $e^{+}p$ 460 GeV & 223 / 204  \\ 
    HERA NC $e^{+}p$ 575 GeV & 223 / 254  \\ 
    HERA NC charm & 49 / 52  \\ 
    HERA NC beauty & 18 / 27  \\ 
    LHCb 7 TeV $B^0$ & 52 / 76  \\ 
    LHCb 7 TeV $B^{+}$ & 129 / 108  \\ 
    LHCb 7 TeV $B^{0}_s$ & 37 / 60  \\ 
    LHCb 7 TeV $D^0$ & 15 / 30  \\ 
    LHCb 7 TeV $D^{+}$ & 19 / 29  \\ 
    LHCb 7 TeV $D^{+}_{s}$ & 14 / 20  \\ 
    LHCb 7 TeV $D^{*+}$ & 16 / 22  \\ 
    LHCb 5 TeV $D^0$ & 60 / 35  \\ 
    LHCb 5 TeV $D^{+}$ & 25 / 35  \\ 
    LHCb 5 TeV $D^{+}_{s}$ & 30 / 29  \\ 
    LHCb 5 TeV $D^{*+}$ & 35 / 30  \\ 
    LHCb 13 TeV $D^0$ & 111 / 60  \\ 
    LHCb 13 TeV $D^{+}$ & 72 / 64  \\ 
    LHCb 13 TeV $D^{+}_{s}$ & 69 / 55  \\ 
    LHCb 13 TeV $D^{*+}$ & 82 / 54  \\ 
    ALICE 7 TeV $D^0$ & 5.1 / 8  \\ 
    ALICE 7 TeV $D^{+}$ & 0.75 / 7  \\ 
    ALICE 7 TeV $D^{*+}$ & 2.3 / 6  \\ 
    ALICE 5 TeV $D^0$ & 6.3 / 10  \\ 
    ALICE 5 TeV $D^{+}$ & 5.8 / 9  \\ 
    ALICE 5 TeV $D^{+}_{s}$ & 2.5 / 4  \\ 
    ALICE 5 TeV $D^{*+}$ & 1.7 / 9  \\ 
    \hline
    Correlated $\chi^2$  & 282  \\ 
    Log penalty $\chi^2$  &  -32  \\ 
    \hline
    Total $\chi^2$ / $n_{dof}$  & 2401 / 1969  \\ 
\end{tabular}
\caption{The global and partial $\chi^2$ values for each data set together with the corresponding number of data points (ndp). The correlated $\chi^2$ and the log penalty $\chi^2$ entries refer to the $\chi^2$ contributions from the correlated uncertainties and from the logarithmic term, respectively, as described in Ref.~\cite{Abramowicz:2015mha}.}
\label{tab:chi}
\end{table}

\begin{table}
    \renewcommand*{\arraystretch}{1.12}
    \tabcolsep1cm
    \centering
\begin{tabular}{ll}
    Parameter & Value \\
    \hline
    $B_g$ & $0.004 \pm 0.053$  \\
    $C_g$ & $6.25 \pm 0.29$  \\
    $Fg$ & $0.068 \pm 0.024$  \\
    $B_{u_v}$ & $0.644 \pm 0.030$  \\
    $C_{u_v}$ & $4.862 \pm 0.076$  \\
    $E_{u_v}$ & $15.8 \pm 2.2$  \\
    $B_{d_v}$ & $0.873 \pm 0.076$  \\
    $C_{d_v}$ & $4.61 \pm 0.35$  \\
    $C_{\overline{U}}$ & $7.36 \pm 0.77$  \\
    $D_{\overline{U}}$ & $10.1 \pm 2.4$  \\
    $A_{\overline{D}}$ & $0.1061 \pm 0.0058$  \\
    $B_{\overline{D}}$ & $-0.1661 \pm 0.0062$  \\
    $C_{\overline{D}}$ & $12.7 \pm 3.0$  \\
\end{tabular}
\caption{The resulting parameters for the PDFs with their fit uncertainties.}
\label{tab:pars}
\end{table}

\begin{figure}
    \centering
    \includegraphics[width=0.49\textwidth]{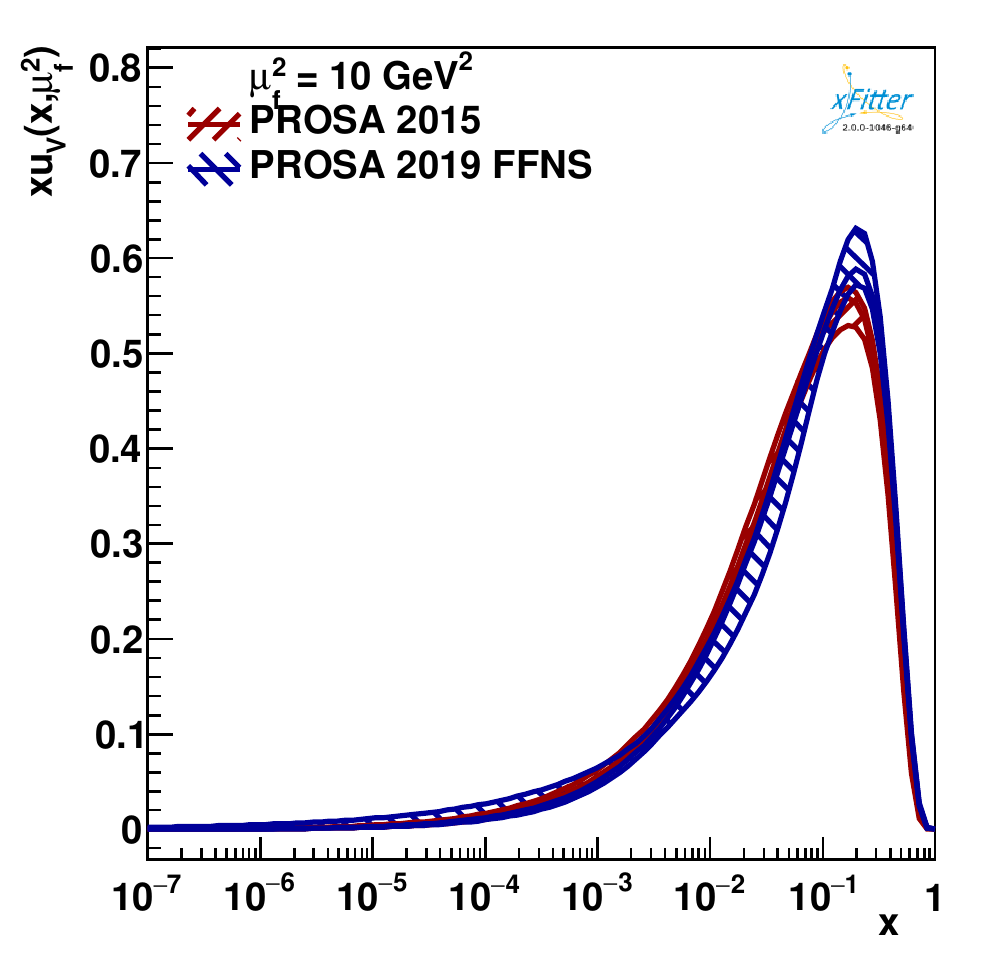}
    \includegraphics[width=0.49\textwidth]{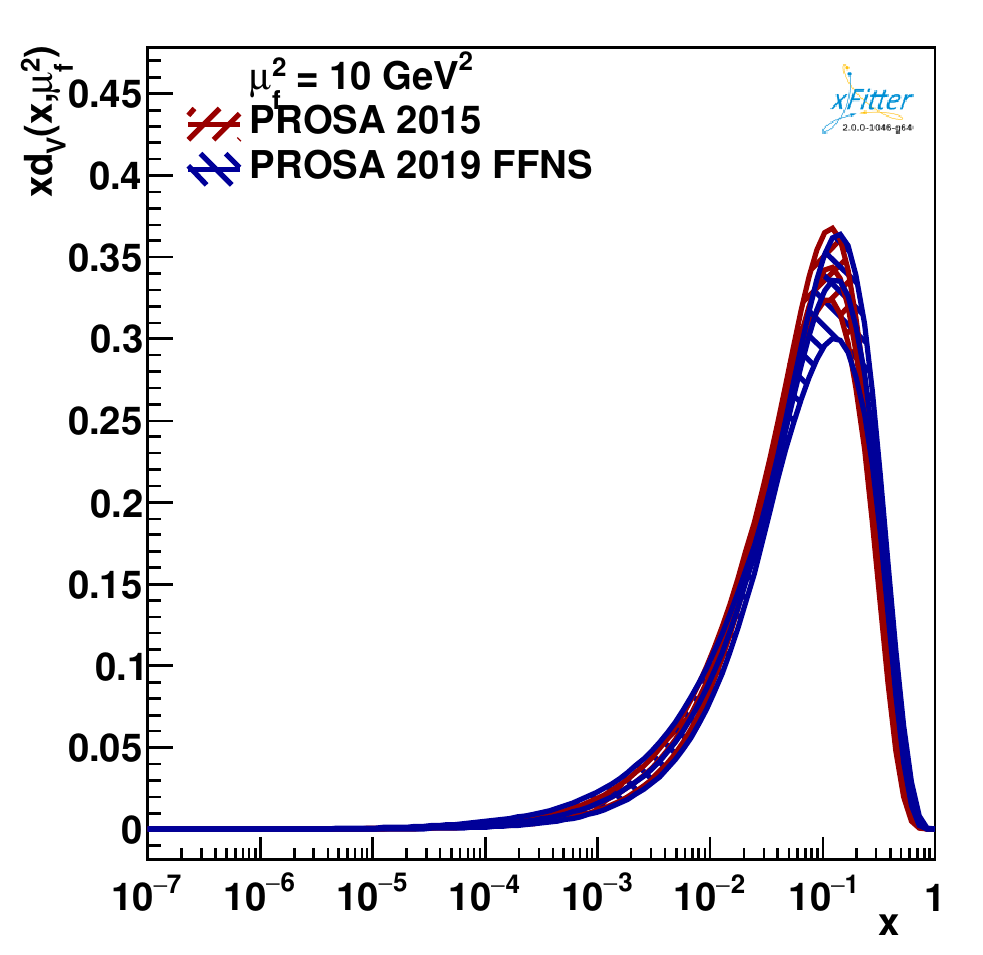}\\
    \includegraphics[width=0.49\textwidth]{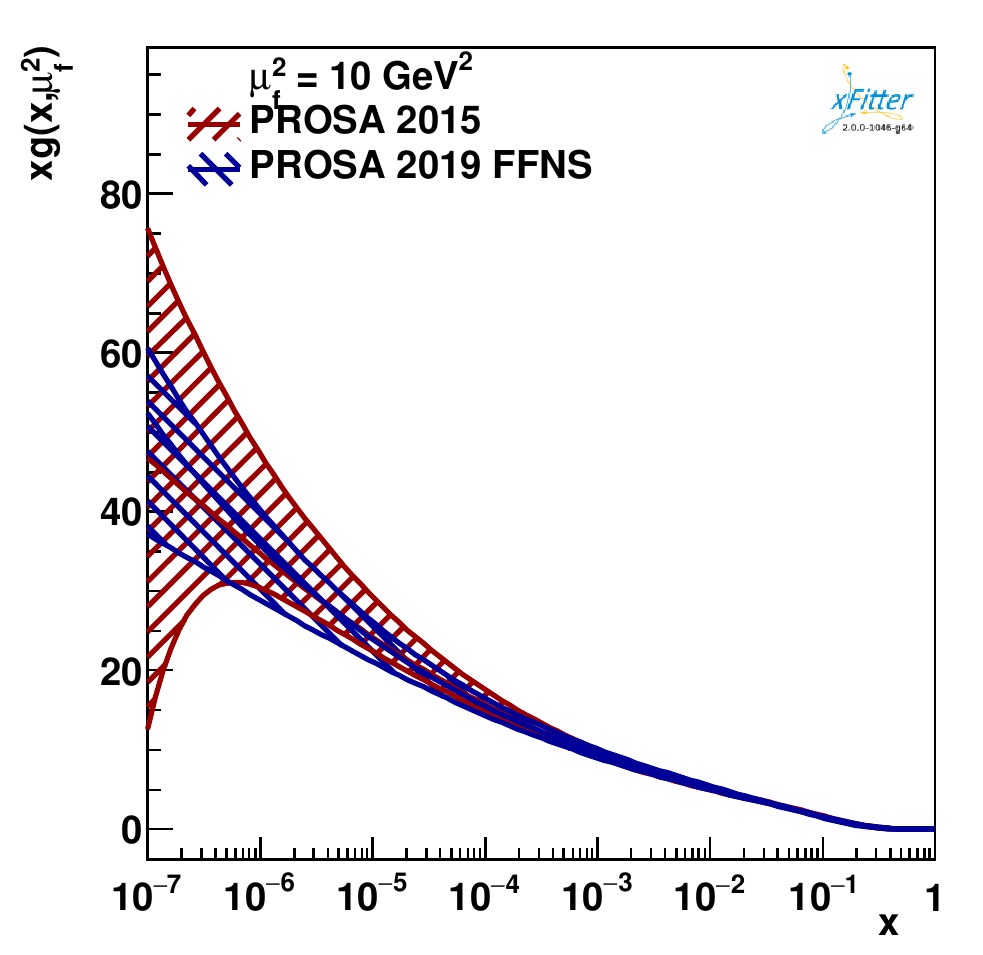}
    \includegraphics[width=0.49\textwidth]{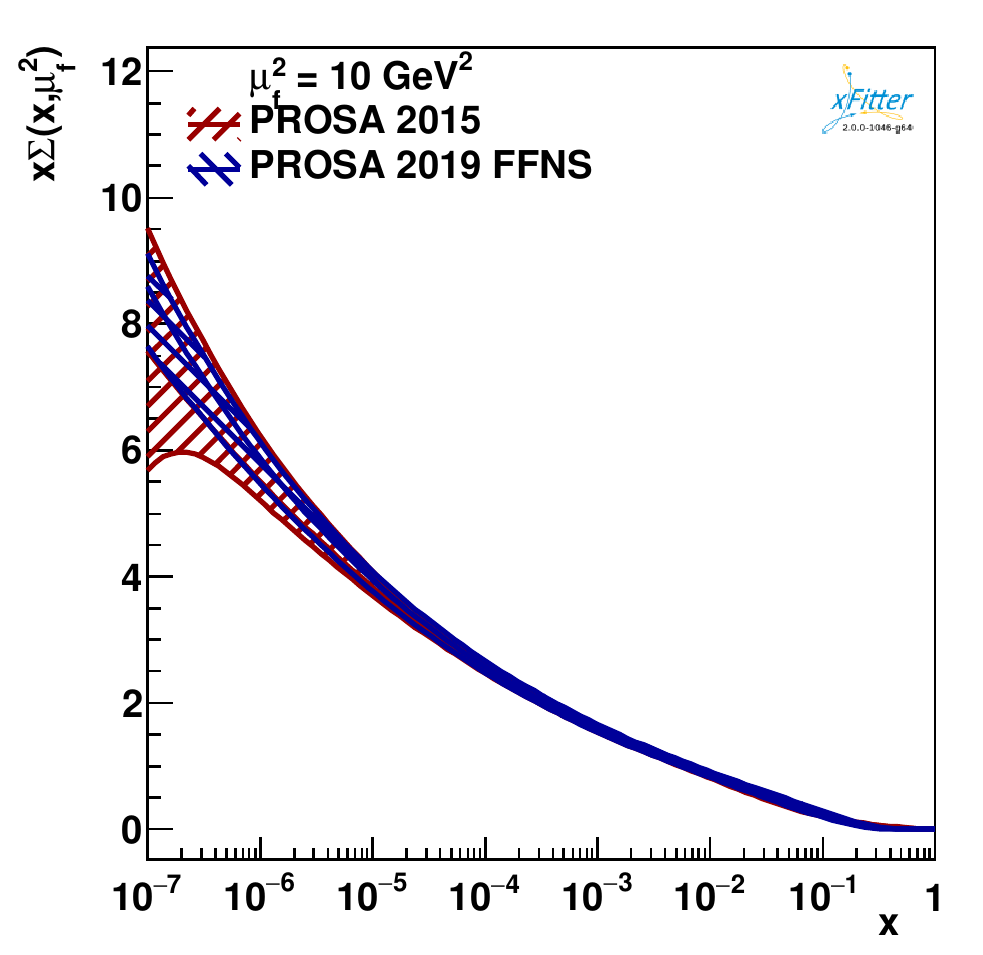}
    \caption{The PROSA 2019 PDF in FFNS with their total uncertainties as a function of $x$ shown at the scale $\mu^2_f=10$ GeV$^2$, compared with the respective distributions from the PROSA 2015 fit.}
    \label{fig:pdfs}
\end{figure}

\begin{figure}
    \centering
    \includegraphics[width=0.49\textwidth]{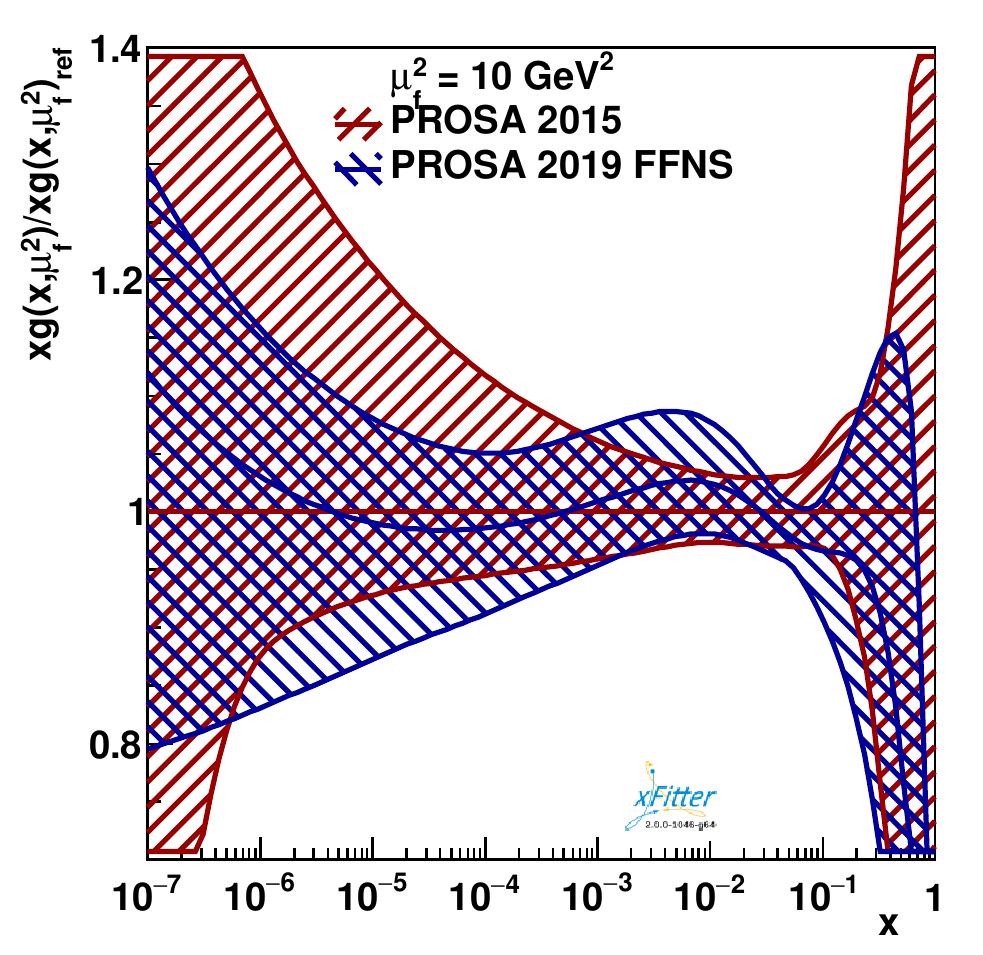}
    \includegraphics[width=0.49\textwidth]{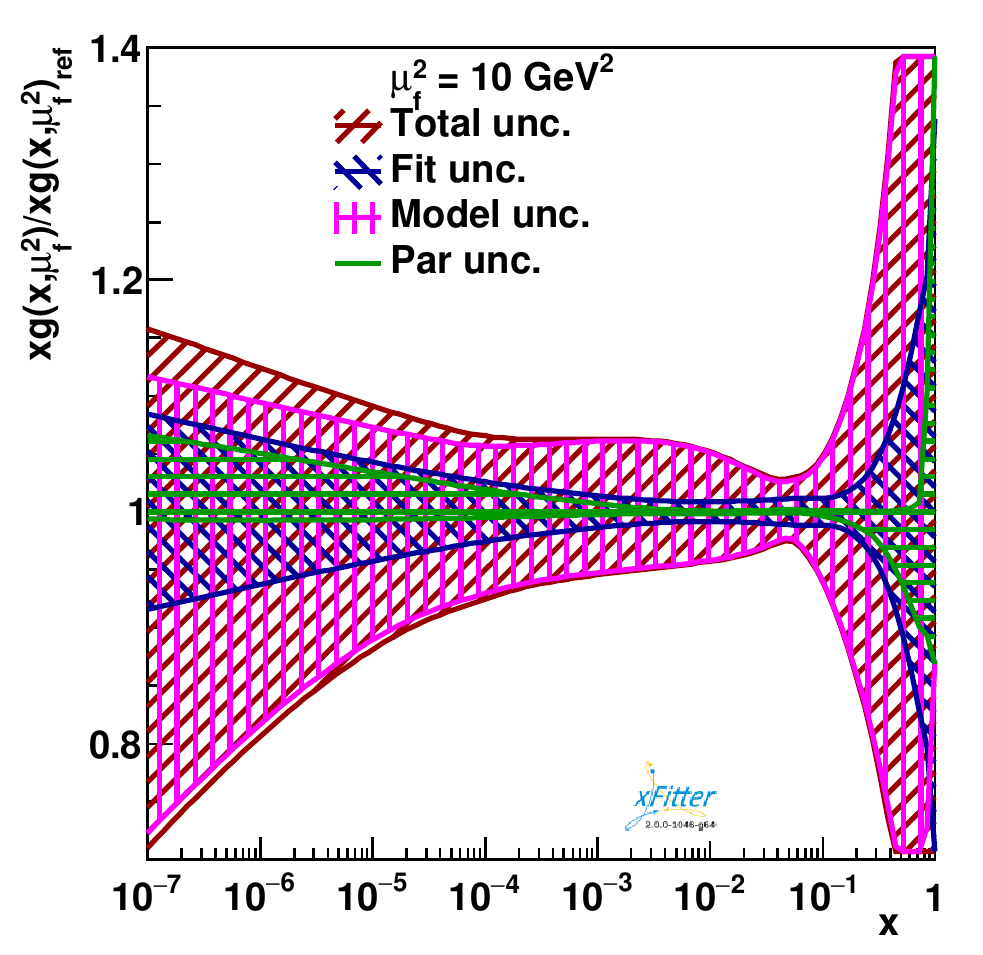}
    \caption{Left panel: the ratio of the gluon distributions of PROSA 2019 FFNS and the PROSA 2015, shown as a function of $x$ at the scale $\mu^2_f=10$~GeV$^2$. Right panel: relative total, fit, model and parametrisation uncertainties for the gluon PROSA 2019 PDF at the scale $\mu^2_f=10$~GeV$^2$.}
    \label{fig:pdfratios}
\end{figure}

The resulting PROSA 2019 PDFs with their total uncertainties at the scale $\mu^2_f=10$~GeV$^2$ are shown in Fig.~\ref{fig:pdfs}. These are compared to the result of the PROSA 2015 fit~\cite{Zenaiev:2015rfa}. In Fig.~\ref{fig:pdfratios} (left), the gluon distribution 
normalised to the one from the PROSA 2015 fit is shown. The two results are in a very good agreement and a significant improvement 
in the precision of the gluon PDF is achieved at $x < 10^{-4}$, as compared to the PROSA 2015 fit. 
{This improvement is attributed mainly to the presence of the new LHCb data in the fit which extends the coverage to lower values of $x$.} 
The valence and sea quark PDFs are in good agreement with the result of the HERAPDF2.0 analysis~\cite{Abramowicz:2015mha} and the 
observed differences in these distributions to the PROSA 2015 analysis are attributed to the update of the DIS measurements~\cite{Aaron:2009aa} used in Ref.~\cite{Zenaiev:2015rfa} to the final combination~\cite{Abramowicz:2015mha} of the HERA data.
{It turns out that once the LHCb data are included in the fit, the inclusion of ALICE data provide no significant additional constraints on the PDFs, but since the ALICE data cover the central range of $y$, the consistent description of these data serves as a non-trivial self-consistency check of the fit using normalised cross sections.}

The relative total, fit, model and parametrisation uncertainties for the gluon PDF are shown in Fig.~\ref{fig:pdfratios} (right). 
The total uncertainties are dominated by the model uncertainties, with the largest contributions arising from the scale 
variations in predictions for heavy-quark hadroproduction. Reduction of these uncertainties would require theoretical calculations at higher order.  The resulting PDFs are available in the LHAPDF format at the PROSA web-page~\cite{prosaweb}.

\subsection{Fit in VFNS}
\label{sec:vfns}

The fit in the VFNS is performed using the APFEL library~\cite{Bertone:2013vaa} interfaced to xFitter.
The theoretical predictions for the HERA data are computed using the FONLL-B scheme~\cite{Forte:2010ta} with the pole charm and beauty quark masses set to $m_c^{\textrm{pole}} = 1.4$ GeV and $m_c^{\textrm{pole}} = 4.5$ GeV respectively.
However, no VFNS calculation for heavy-quark $pp$ hadroproduction is interfaced to public QCD analysis tools like xFitter.
To use the MNR calculations with the VFNS, the functionality of the APFEL library is exploited, allowing to choose arbitrary heavy-quark matching thresholds~\cite{Bertone:2017ehk}. These thresholds are set as:
\begin{equation}
\begin{aligned}
\mu_c &= 4.5m_c^{\textrm{pole}} = 6.3~\textrm{GeV},\\
\mu_b &= 4.5m_b^{\textrm{pole}} =  20.25~\textrm{GeV}.
\label{eq:thr}
\end{aligned}
\end{equation}
The kinematic requirements $p_T < 5$ GeV and $p_T < 16$ GeV are imposed on the LHC charm and beauty data, respectively, to ensure that not more than 3 (4) flavours are considered when calculating predictions for charm (beauty) data when choosing $\mu_r = \mu_f = \sqrt{Q^2 + 4m_Q(m_Q)^2}$.
The strong coupling constant is set to $\alpha_s^{n_f = 5}(M_Z) = 0.118$~\cite{Tanabashi:2018oca}, while all other settings are the same as in the FFNS fit.
The specific matching thresholds in Eq.~(\ref{eq:thr}) are chosen to ensure that a sufficient amount of the LHC charm and beauty data is still included in the fit.
The choice of the matching thresholds is arbitrary and coincides with the renormalisation scheme choice of Ref.~\cite{Bertone:2017ehk}. 
The results are proven to remain stable under variations of $3.1 \le \mu_Q/m_Q^{\textrm{pole}} \le 6$, whereby the $p_T$  cuts in the charm and beauty cross-section measurements of the LHC are modified accordingly. 

In the VFNS variant of the PDF fit, $\chi^2 = 2114$ is obtained for $n_{dof} = 1714$, indicating a similar quality of data description as compared to the fit in the FFNS {(consistent with the observations in Refs.~\cite{Abramowicz:2015mha,H1:2018flt}).} The resulting PDFs are available in the LHAPDF format at the PROSA web-site~\cite{prosaweb}. No PDF uncertainties are provided with this set.

The performance of the PROSA 2019 VFNS PDFs is tested by computing predictions for the inclusive and multi-jet production in DIS~\cite{Chekanov:2002be,Chekanov:2006xr,Abramowicz:2010cka,Aktas:2007aa,Aaron:2010ac} 
and jet~\cite{Chatrchyan:2012bja} and top quark-antiquark production~\cite{Sirunyan:2017azo,Sirunyan:2019zvx} in $pp$ collisions. The results collected at the PROSA web-site~\cite{prosaweb} are found to be similar to those using HERAPDF2.0 PDF.

\begin{figure}
	\centering
	\includegraphics[width=0.49\textwidth]{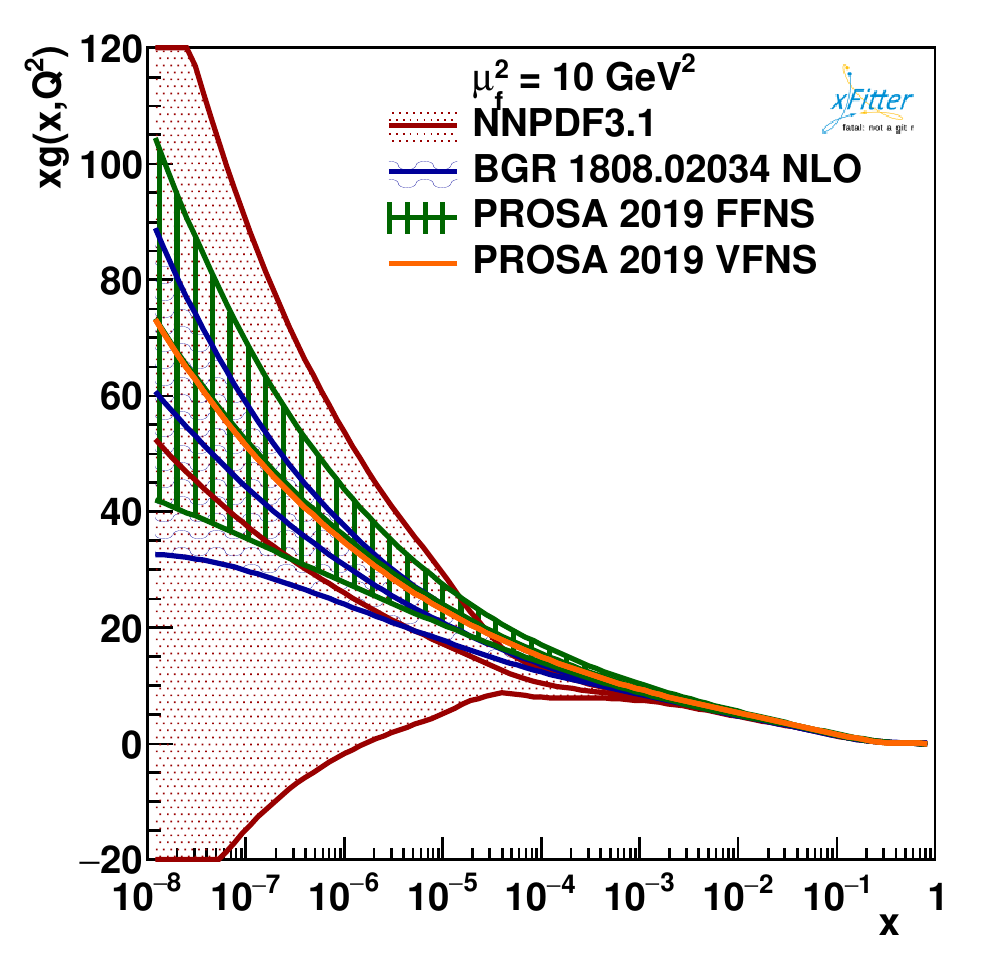}
	\caption{The gluon distributions of PROSA 2019 FFNS and VFNS fits compared with the NNPDF3.1~\cite{Ball:2017nwa} PDFs and with the PDFs from Ref.~\cite{Bertone:2018dse} obtained using NLO calculations at the scale $\mu^2_f=10$~GeV$^2$.}
	\label{fig:vfns}
\end{figure}

{ In Fig.~\ref{fig:vfns} we compare the gluon distribution obtained in our fits in FFNS and VFNS with the NNPDF3.1 PDFs~\cite{Ball:2017nwa} and with the PDFs from Ref.~\cite{Bertone:2018dse} obtained using NLO calculations which also exploit the LHCb measurements of D-meson cross-sections at 5, 7 and 13 TeV. The different fits are in good agreement.}

\section{Predictions for prompt atmospheric-neutrino fluxes}
\label{sec:astro}
Various applications in high-energy astroparticle physics could benefit from accurate PDFs in the low-$x$ region. One of the most interesting cases is the evaluation of the prompt flux of atmospheric neutrinos,
originating from the semileptonic decays of heavy-flavoured hadrons produced in the interactions of cosmic rays (CR) with nuclei in the atmosphere. The prompt atmospheric-neutrino flux represents a relevant background for searches of highly energetic cosmic neutrinos, which are supposed to be produced in the vicinity of far astrophysical sources and in the Galactic Plane~\cite{Gaisser:2016uoy}.  
Such searches are conducted at Very Large Volume Neutrino Telescopes such as ANTARES~\cite{Collaboration:2011nsa}, IceCube~\cite{Gaisser:2014foa} and KM3NeT~\cite{Adrian-Martinez:2016fdl}, which register and analyse the features of the track and cascade events induced by the charged-current and neutral-current weak interactions of the impinging neutrinos with the water/ice nuclei. To date, no direct measurement of the prompt atmospheric-neutrino flux is available. Therefore, the most precise theoretical predictions for these fluxes are needed for the reliable interpretation of the experimental data in order to disentangle the cosmic neutrino component from the atmospheric background~\cite{Mascaretti:2019uqn}.

In this paper, the predictions for the prompt atmospheric-neutrino fluxes are calculated, in general following the method detailed in Ref.~\cite{Garzelli:2016xmx}. It is assumed, that $pA$ and $AA$ interactions leading to charm production can be described in terms of $pp$ interactions (superposition model) in pQCD. For the proton structure description, the PROSA 2019 PDF fit is used among other PDFs.
Production and decay of the $D^\pm$, $D^0$, $\bar{D}^0$, $D_s^\pm$, $\Lambda_c^\pm$ in the atmosphere is considered dominant, 
since the contribution of other charmed hadrons, as well as b-flavoured hadrons, amounts to 5-15\% of the dominant one~\cite{Bhattacharya:2016jce}. In the computation of charmed-hadron production cross sections, the renormalisation and factorisation scales are chosen as $\mu_R$ = $\mu_F$ = $\mu_0$ = $\sqrt{p_T^2 + 4 m_c^2}$, consistent with the scale choice adopted in the theory predictions of $D$- and $B$-meson production at LHCb and ALICE used in the PDF fit. Note that this scale choice differs from the one of Ref.~\cite{Garzelli:2016xmx} (PROSA 2015), where $\mu_R$ = $\mu_F$ = $\sqrt{p_T^2 + m_c^2}$ was used, consistent with~\cite{Zenaiev:2015rfa}. While the difference between the two scale choices reduces with increasing $p_T$, at low $p_T$ the present scale choice is motivated by faster convergence of the perturbative series to NNLO for the total $pp$~$\rightarrow$~$c\bar{c} + X$ cross section at the LHC energies, as reported in Ref.~\cite{Garzelli:2015psa}. 

In the present work, the central value of the pole mass of the charm quark, $m_c^{pole}$ = 1.43~GeV is used, corresponding to $m_c(m_c) = 1.23$ GeV in the PDF fit (see Table 2), as obtained using 1\nobreakdash-loop conversion. It is worthwhile to note that this value is somewhat larger then the one\footnote{In Ref.~\cite{Garzelli:2016xmx}, the $m_c$=1.4 GeV was used instead of the $m_c$=1.25 GeV obtained in the PDF fit of Ref.~\cite{Zenaiev:2015rfa}.} used in the PROSA 2015 computation.  
The uncertainty due to the choice of the charm quark mass is evaluated by varying the pole mass by~$\pm$~0.15~GeV around the central value.

The PDF uncertainties are evaluated using the respective uncertainty eigenvectors, provided. 
The uncertainty related to the choice of the scales is evaluated considering the envelope of the resulting cross sections for the assumptions ($\mu_R$, $\mu_F$) = \{(1,~1), (0.5,~0.5), (2,~2), (1,~2), (2, 1), (1, 0.5), (0.5, 1)\} ($\mu_0$, $\mu_0$). 
  
\begin{figure}
    \centering
    \includegraphics[width=0.49\textwidth]{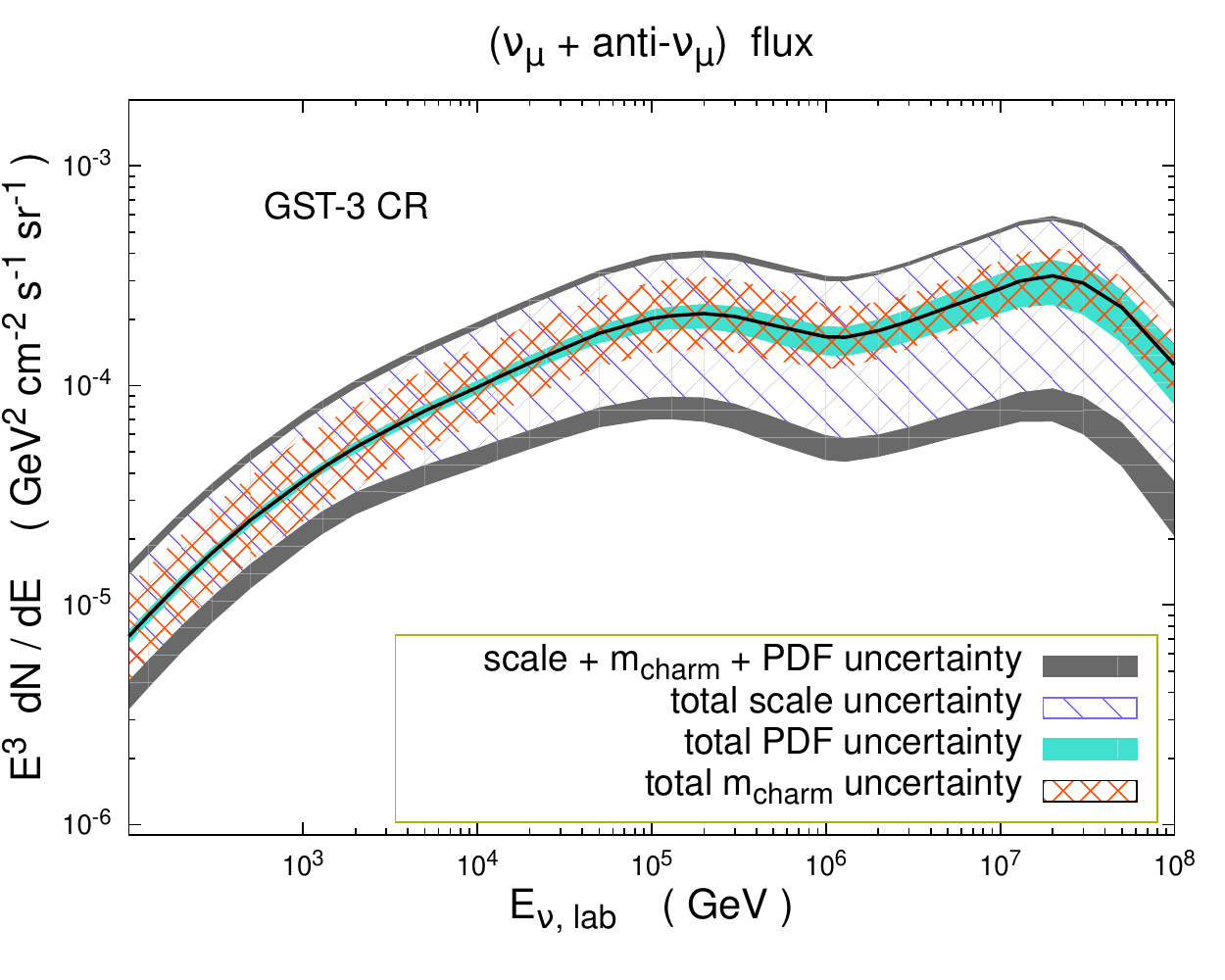}
    \includegraphics[width=0.49\textwidth]{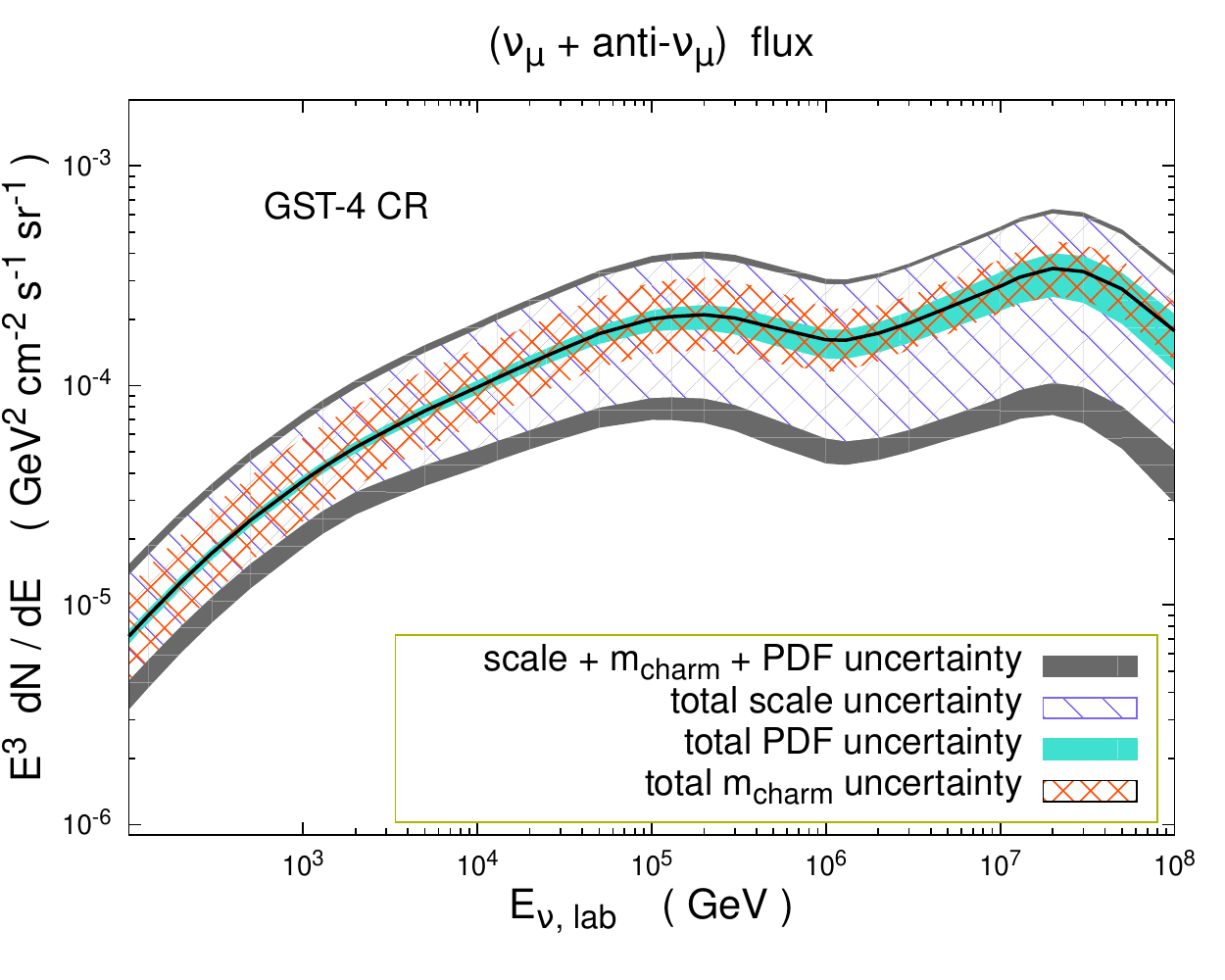}
    \includegraphics[width=0.49\textwidth]{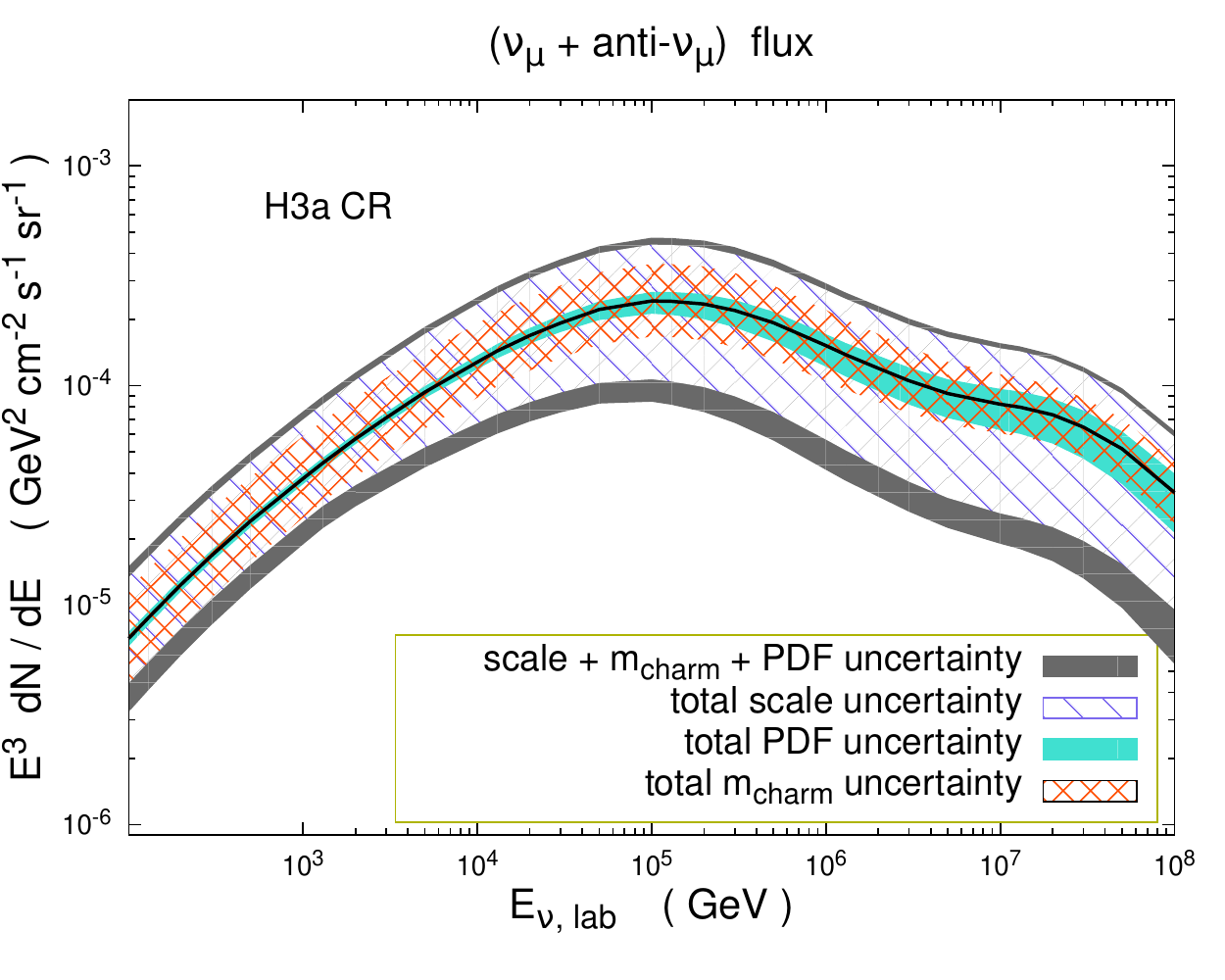}
    \includegraphics[width=0.49\textwidth]{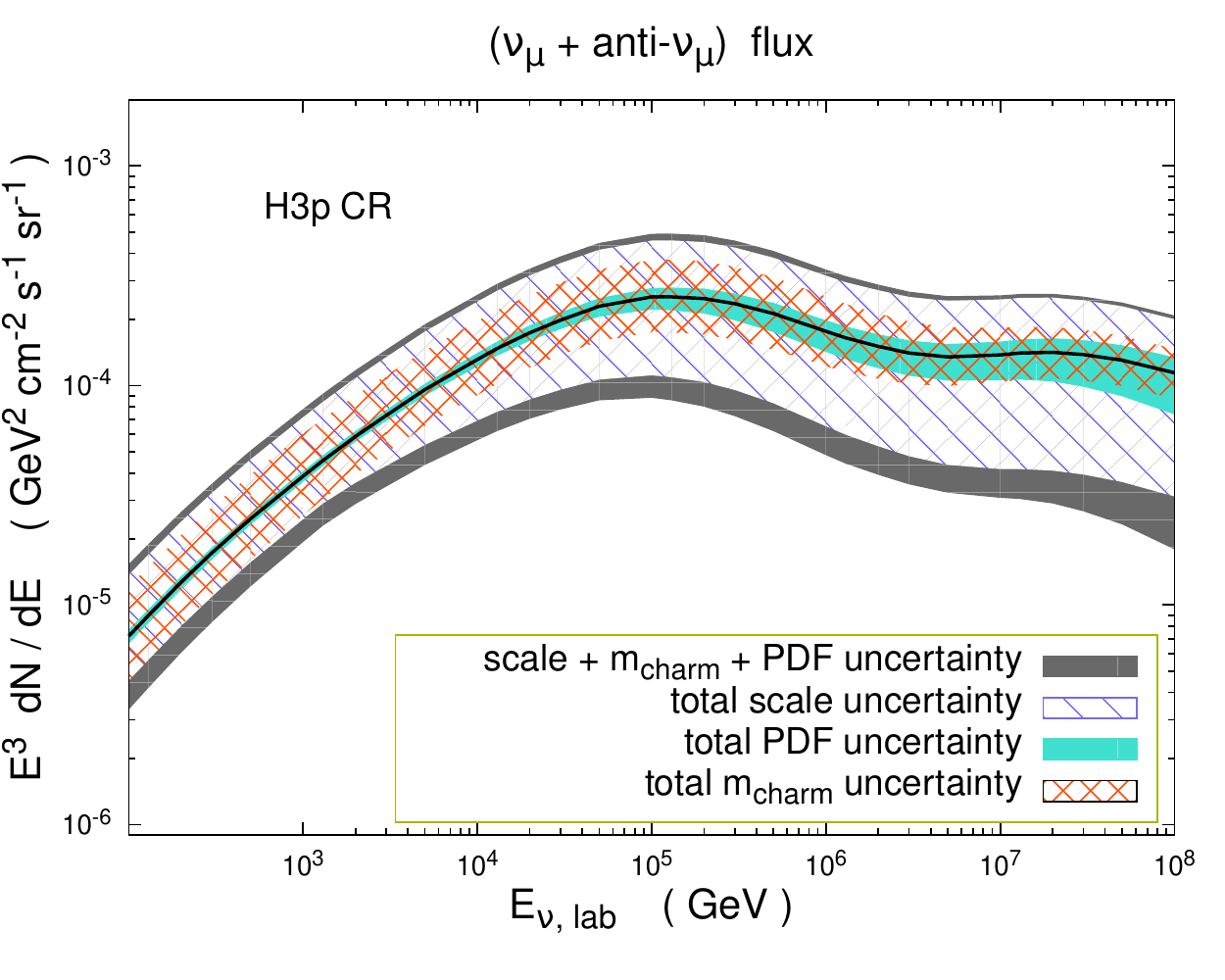}
    \includegraphics[width=0.49\textwidth]{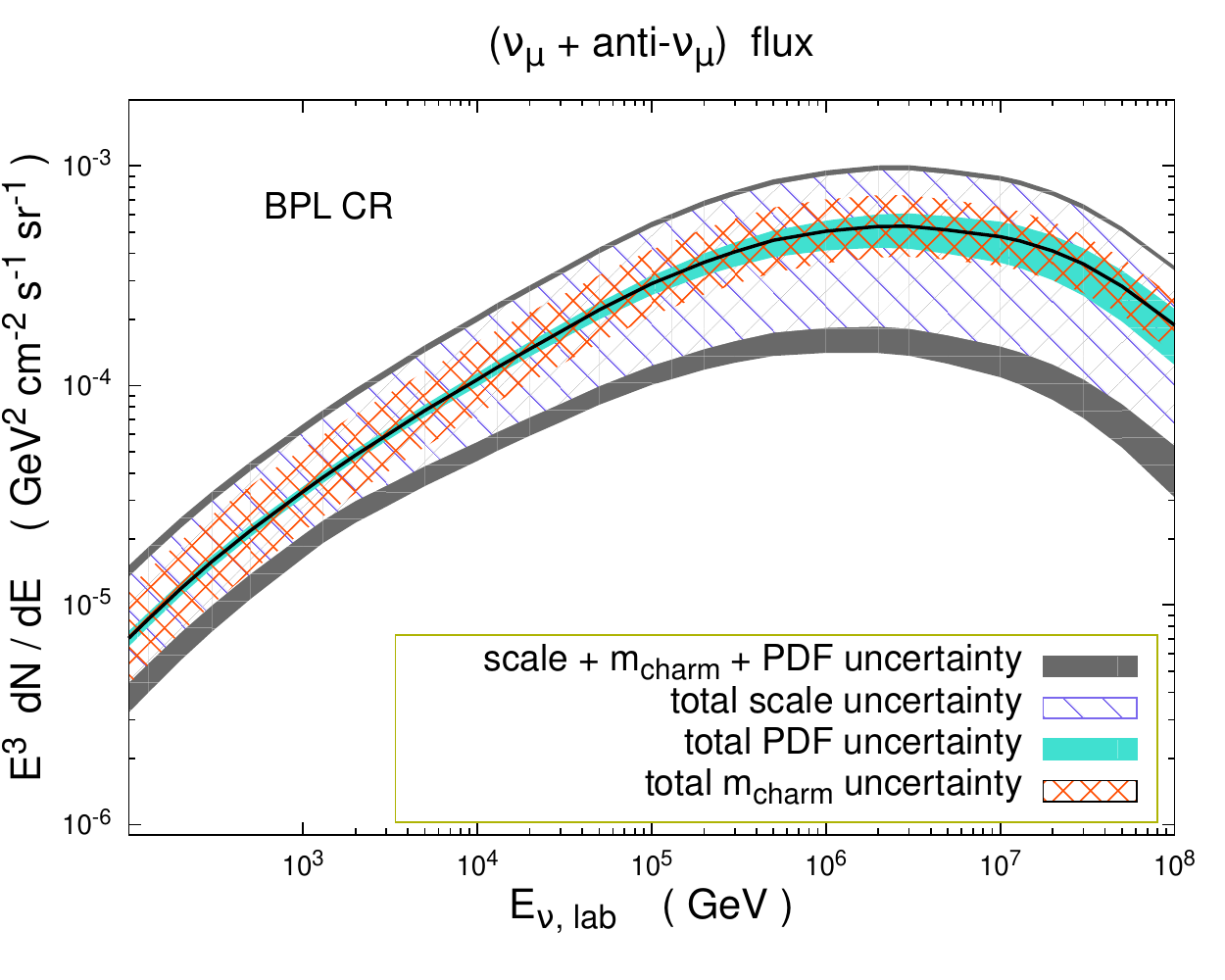}
    \includegraphics[width=0.49\textwidth]{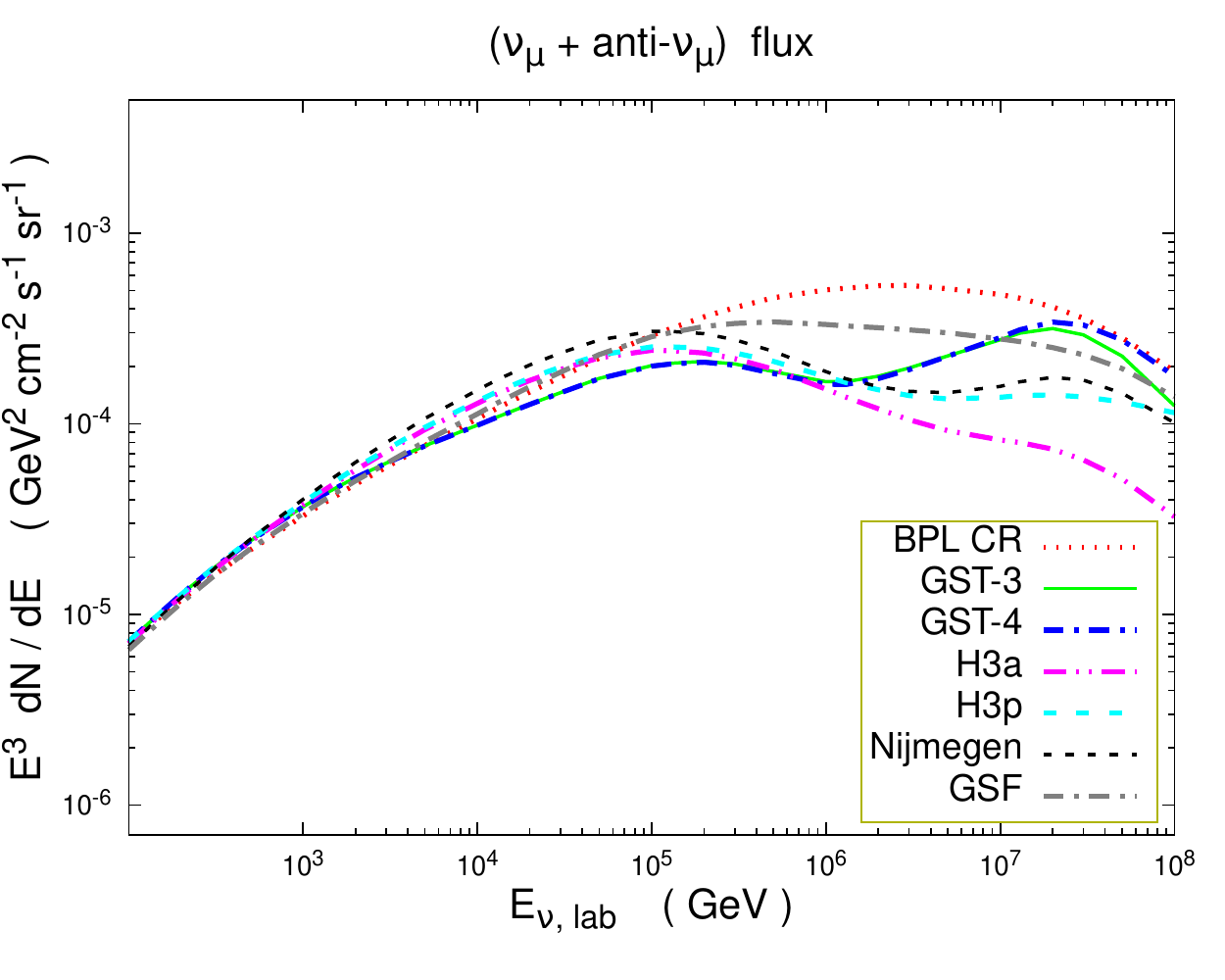}
  \caption{\label{fig1prompt}  Predictions for the prompt atmospheric-neutrino fluxes and their uncertainties related to scale variation, charm mass and PDF uncertainties. Each of the first five panels refers to a different CR primary all-nucleon spectrum (GST-3, GST-4, H3a, H3p~\cite{Gaisser:2013bla, Gaisser:2011cc},  and Broken-Power-Law (BPL)~\cite{Gaisser:2016uoy}, respectively), chosen among those most widely used in literature. In the bottom panel on the right-hand side the central predictions using all these primary all-nucleon spectra are compared among each other, and with those obtained with the more recently introduced Nijmegen~\cite{Thoudam:2016syr} and Global Spline Fit (GSF)~\cite{Dembinski:2017zsh,Schroeder:2019agg} all-nucleon spectra. At the highest $E_{\nu\, lab}$, the largest predictions (GST-4 and BPL) are seven time larger than the smallest one (H3a).}
\end{figure}

The predictions for the prompt ($\nu_\mu$ + $\bar{\nu}_\mu$) fluxes using PROSA 2019 PDFs are presented in Fig.~\ref{fig1prompt}. Those are obtained by using different hypotheses for the primary CR all-nucleon flux~\cite{Gaisser:2011cc,Gaisser:2013bla}, which are derived from the measured CR all-particle spectrum~\cite{Kachelriess:2019oqu}, under specific assumptions for the CR composition. In particular, these assumptions concern the proton and nuclear groups included in the derivation of the spectra, their spectral indices, their rigidity, the number of populations of galactic origin and their sources, the presence or not of an additional proton population of extragalactic origin, as detailed in the aforementioned references.   

The QCD uncertainties in the resulting prompt ($\nu_\mu$ + $\bar{\nu}_\mu$) fluxes encompass the uncertainties in the charm quark mass, PDF and those related to the scale choice, with the latter being the dominant uncertainty. The quoted scale uncertainty bands are obtained at fixed PDF, i.e. they do not include the contribution related to scale variation in the PDF fit. The effect of varying the ($\mu_R$, $\mu_F$) scales when comparing theoretical predictions with experimental data in the PDF fit process, according to the method detailed in Section 4, is instead accounted for in the PDF uncertainty band.  
As expected, the uncertainty on prompt neutrino fluxes due to the variation of the charm quark mass value in the constant interval around its central constant value, decreases with energy: at small $E_{\nu,\,lab}$ this uncertainty dominates over the PDF uncertainty, whereas at $E_{\nu,\,lab} \sim 10^7$~-~$10^8$~GeV, both uncertainty contributions become similar. 
At high $E_{\nu,\,lab}$, the PDF uncertainties are reduced with respect to those of the PROSA 2015 computation. 

\begin{figure}
\centering
    \includegraphics[width=0.49\textwidth]{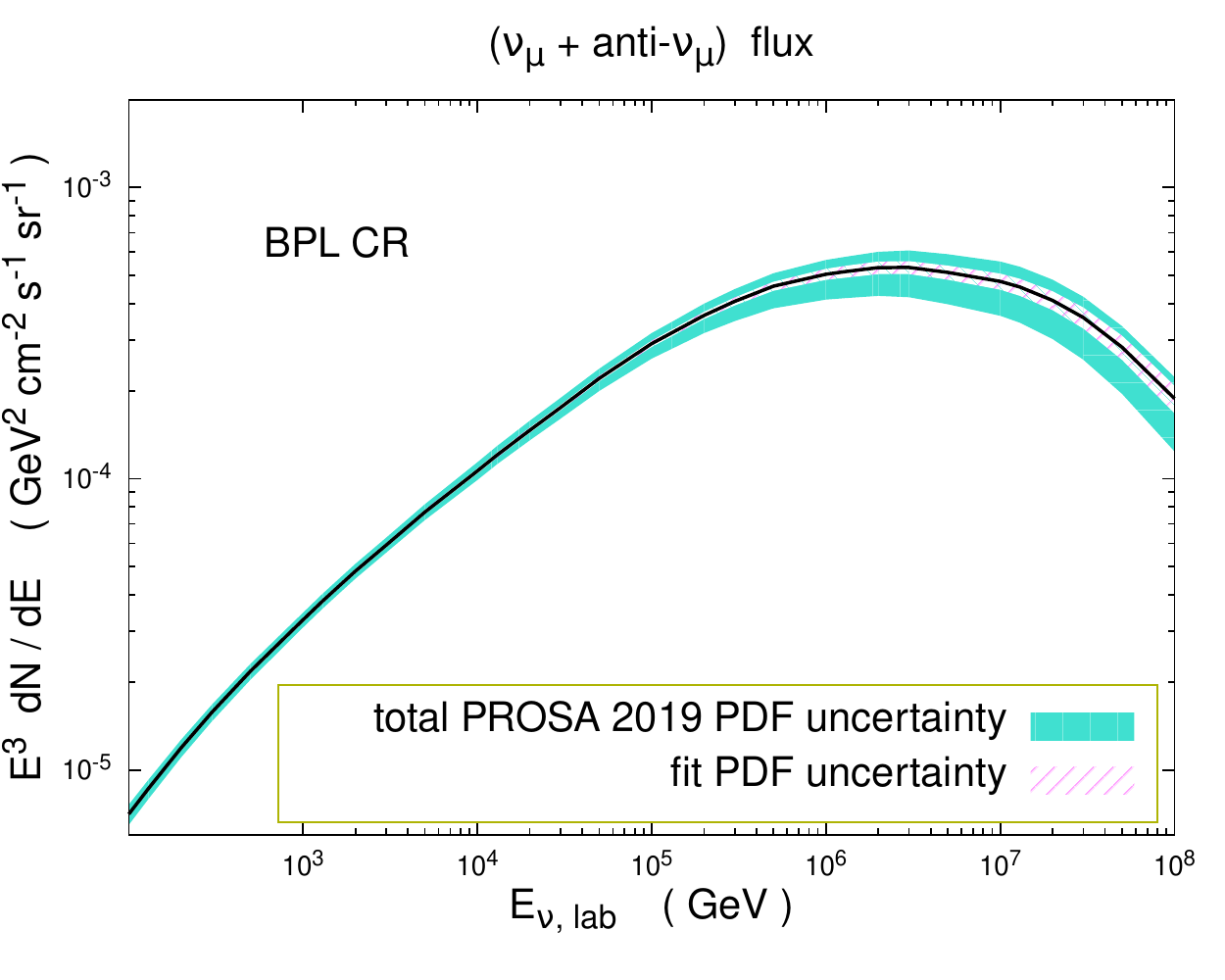}
    \includegraphics[width=0.49\textwidth]{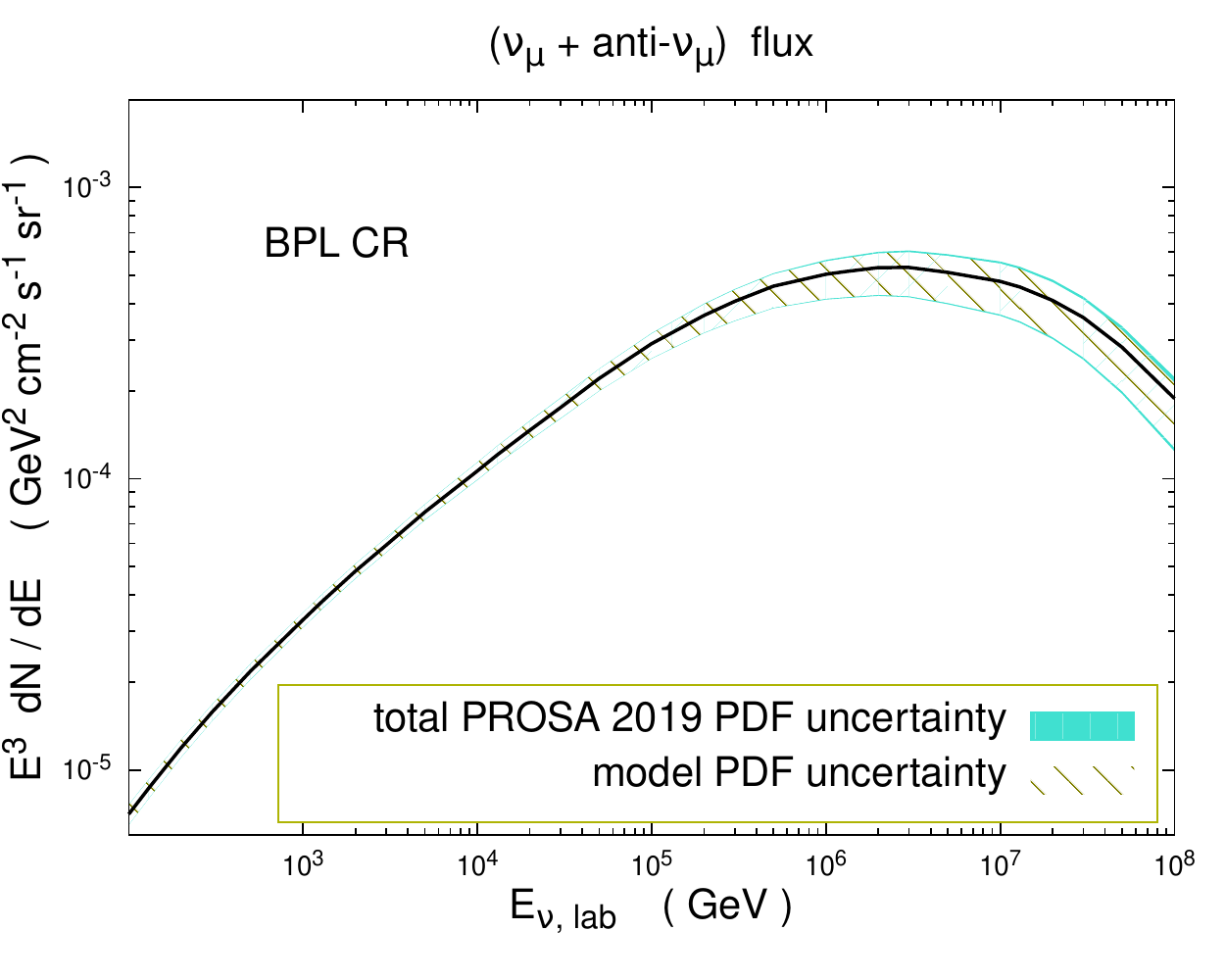}
    \includegraphics[width=0.49\textwidth]{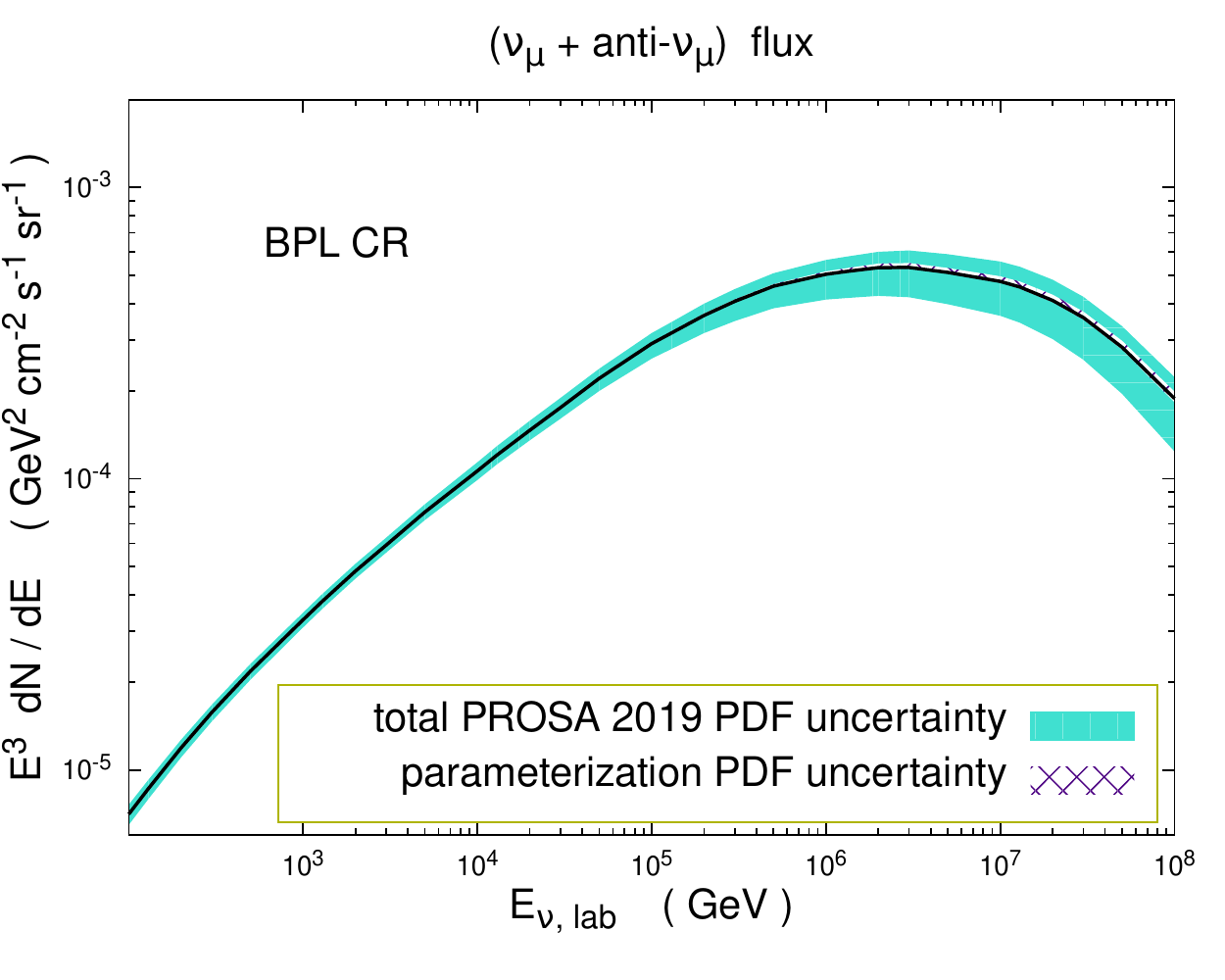}
\caption{\label{fig2prompt} Contributions of the PDF fit (top left), model (top right) and parametrisation (bottom) uncertainties to the total uncertainties in the prediction for the prompt atmospheric-neutrino flux using the BPL primary CR all-nucleon spectrum.}  
\end{figure}

The different contributions of the PROSA 2019 PDF uncertainty in the flux prediction are shown in Fig.~\ref{fig2prompt}. All the uncertainties increase with increasing $E_{\nu,\,lab}$, which corresponds to the decreasing $x$ of the target parton, probed. 

Prompt neutrinos with energy $E_{\nu,\,lab}$ are mostly produced by air collisions of CR protons with laboratory energies (10 - 100) times larger. Therefore, neutrinos with energies of some PeV, i.e. the most energetic neutrinos seen so far by IceCube, are mostly obtained by collisions up to the LHC centre-of-mass energies. On the other hand, neutrinos with higher energies can be the result of collisions at energies not yet probed at accelerators. It is worthwhile to note that the PDF uncertainties for $10^6$ $<$ $E_{\nu,\,lab}$ $<$ $10^8$ GeV are calculated assuming the PDFs can be extrapolated to $x$-values lower than the kinematic reach of the data used in the PDF fit, $x \approx 10^{-6}$. For an $x$-range lower than $10^{-6}$, there are no measurements probing this region to date.\footnote{We recall that the kinematic formula relating the projectile/target parton $x$ with the $p_T$ and $y$ of a produced heavy-quark with mass $m_Q$ in $pp \rightarrow Q\bar{Q}$ collisions at a laboratory energy $E_p$ is $x$ = $\frac{\sqrt{p_T^2 + m_Q^2}}{E_p}{\rm e}^{\pm y}$.}. However, the computation of the prompt neutrino flux at the highest $E_{\nu,\, lab}$ energies involves a non-negligible contribution from initial state partons with $x$ lower than $10^{-6}$, as shown in Fig.~4 of~\cite{Goncalves:2017lvq}.      
The agreement of the results based on the PROSA 2019 and the PROSA 2015 PDF sets can be considered as a consistency test of the extrapolation procedure, assuming no New Physics contribution in the probed $x$-range. Furthermore, at the neutrino energies of $E_{\nu,\,lab}~\gtrsim~10^5$ GeV, the assumptions on the CR composition become very important (see Fig.~\ref{fig1prompt}, bottom right-hand plot), having an impact on both the shape and the normalisation for the prompt atmospheric-neutrino flux. In particular, at the highest $E_{\nu,\, lab}$, corresponding to the lowest $x$ values, the spread between central predictions obtained using as input different CR primary all-nucleon spectra amounts to a factor of $\mathcal{O}$(5-10), that is much larger than the extrapolated PDF uncertainty.  

\begin{figure}
\centering
    \includegraphics[width=0.62\textwidth]{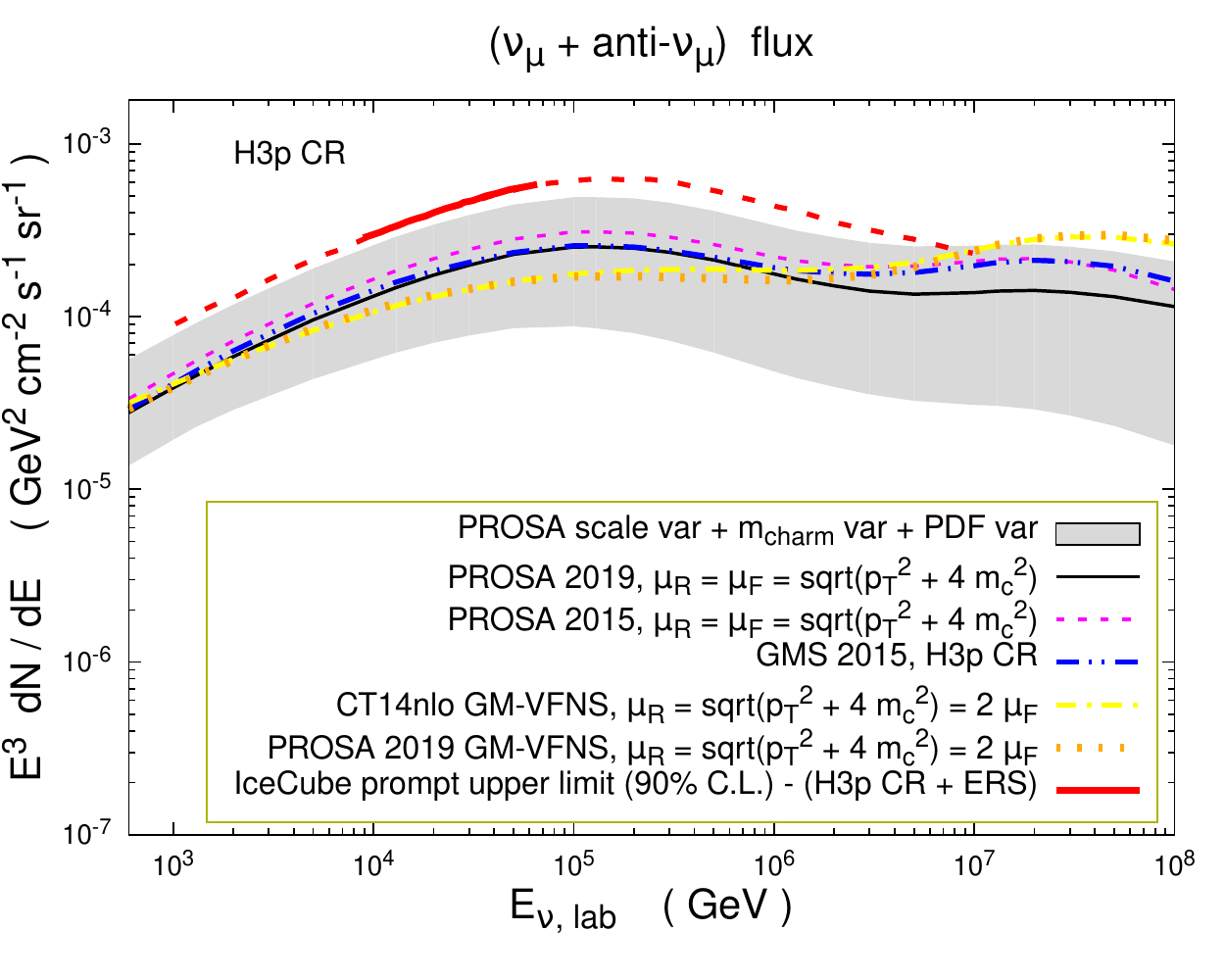}
  \caption{\label{fig3prompt} 
Predictions for prompt neutrino fluxes from this paper as compared to other predictions previously obtained by some members of our group. Predictions obtained with the PROSA PDF fit of Ref.~\cite{Zenaiev:2015rfa}, using the same $\mu_R$ = $\mu_F$ scale, but a slightly different charm mass value ($m_c$ = 1.4 GeV) are shown by dotted (pink) lines; the predictions of Ref.~\cite{Garzelli:2015psa}, using the ABM11 PDF fit~\cite{Alekhin:2012ig}, are shown by double-dotted-dashed (blue) lines. Finally the predictions using the VFNS version of the PDF fit described in this paper, in association with a GM-VFNS calculation of charm hadroproduction~\cite{Benzke:2017yjn}, are also reported and compared to those of Ref.~\cite{Benzke:2017yjn} itself that made use of the CT14nlo PDF fit~\cite{Dulat:2015mca}. 
The experimental upper limit on prompt ($\nu_\mu$ + $\bar{\nu}_\mu$) flux, extracted from IceCube in the analysis of Ref.~\cite{Aartsen:2016xlq}, is also reported. The solid red line is the limit inferred by the available IceCube data, whereas the dashed line is its extrapolation to different $E_\nu$, still computed by the IceCube collaboration.}
\end{figure}

In Fig.~\ref{fig3prompt}, predictions for prompt atmospheric-neutrino fluxes using different descriptions of the proton structure, are compared among each other. The predictions using FFNS PROSA 2019, PROSA 2015 and ABM11 PDFs with corresponding $\alpha_S(M_Z)$, have been obtained using as a basis matrix elements for $c\bar{c}$ hadroproduction at NLO in FFNS ($N_f$=3), matched, according to the POWHEG formalism~\cite{Nason:2004rx, Frixione:2007nw}, to the {\texttt{PYTHIA8}} parton shower and hadronisation algorithms\footnote{In this work the PROSA 2019 PDFs are used for the evaluation of the fixed-order cross-sections but not in the {\texttt{PYTHIA}} Shower Monte Carlo, where PDFs consistent with the tunes already available are instead used. Changing PDFs in the Shower Monte Carlo code would require its re-tuning.}~\cite{Sjostrand:2014zea}. Each $pp$ collision can produce more than two $D$-hadrons, because charm quarks can be produced both in the hard-scattering and during the parton shower processes. 
The predictions using PROSA 2019 at high energies are somewhat lower than those using PROSA 2015 and ABM11 PDFs.
In the same Figure, the predictions obtained in the general-mass VFNS framework of Ref.~\cite{Benzke:2017yjn} (GM-VFNS), using as input 
VFNS PDFs (CT14nlo and the PROSA 2019 VFNS) are shown. The NLO QCD corrections are included in the partonic cross section, whereas the transition from partons to hadrons is described by fragmentation functions evolving with the factorisation scale~\cite{Kneesch:2007ey}, 
a procedure which resums logarithms of $p_T/m_c$ at next-to-leading-logarithmic accuracy. Both central predictions using the GM-VFNS shown in the plot are compatible among each other, but show shape differences with respect to the FFNS ones. Part of these differences are related to the different treatment of the transition of partons into hadrons (parton shower + hadronisation on the one hand, vs. fragmentation functions on the other hand). Also, a different factorisation scale is used in the GM-VFNS predictions~\footnote{Motivations for the specific $\mu_F$ = $\mu_R$/2 choice adopted in the GM-VFNS computation are reported in Ref.~\cite{Benzke:2017yjn}.}. For comparison, the upper limit on the prompt neutrino flux obtained in the IceCube analysis~\cite{Aartsen:2016xlq} of up-going muons 
from the northern hemisphere is also shown and is well described by the predictions. 

\begin{figure}
\centering
    \includegraphics[width=0.7\textwidth]{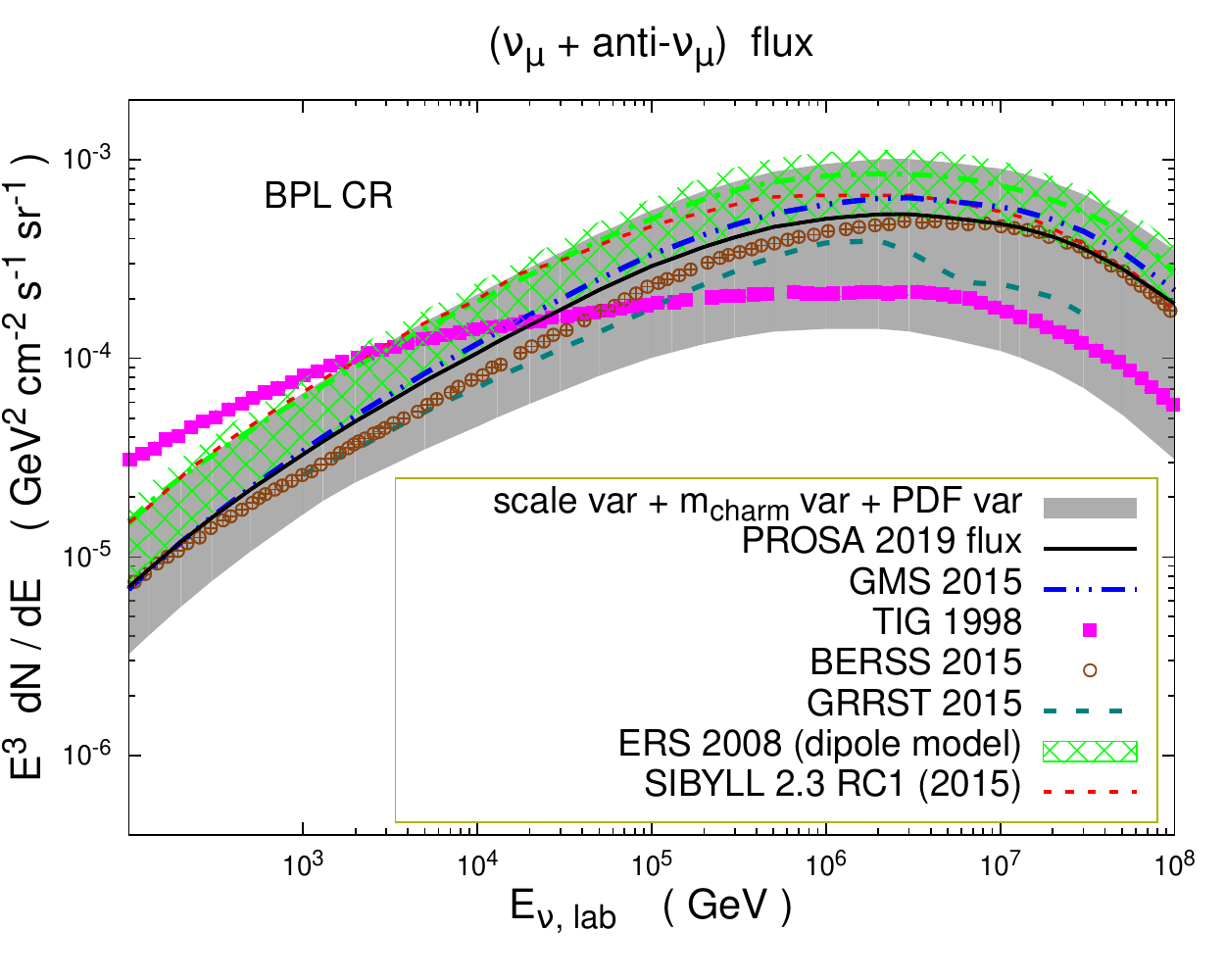}
  \caption{\label{fig4prompt} Predictions for prompt atmospheric-neutrino fluxes obtained in the presented analysis, compared to those by other authors~\cite{Garzelli:2015psa, Gondolo:1995fq, Enberg:2008te, Bhattacharya:2015jpa, Gauld:2015kvh, Fedynitch:2015zma}, presented by lines of different style. The primary CR all-nucleon flux BPL is used in all the predictions.}
\end{figure}

The various predictions shown in Fig.~\ref{fig3prompt} were all obtained under identical assumptions for all inputs used in the solution of the cascade equations for the evaluation of prompt neutrino fluxes, except for the explained differences in the evaluation of $D$-hadron production. On the other hand, in Fig.~\ref{fig4prompt} the presented flux prediction using the PROSA 2019 PDFs is compared to those obtained by other groups. The BPL primary CR all-nucleon spectrum~\cite{Gaisser:2016uoy} is used as input for this comparison because of its very simple form which has allowed an easy incorporation of this spectrum in the computation of many different authors. Although the general methodology for the calculation of prompt neutrino fluxes is the same, the calculation by different authors are obtained in a completely independent way and, thus, at least in principle, might differ in many respects, not only related to the methodology for the computation of charm hadroproduction, but even for other assumptions in the solution of cascade equations (e.g. the details of the atmospheric model, of the $p$-Air total inelastic cross-section and of the proton and hadron regeneration processes~\cite{Garzelli:2015psa}). Notwithstanding these possible further sources of discrepancies, we observe that our predictions turn out to be consistent with those by other authors, within uncertainties. Due to this similarity, the experimental collaborations, in many of their works, limit themselves to consider as input only very few (if not only one) of the theoretical predictions available. The result of Ref.~\cite{Gondolo:1995fq} shows the largest differences with the presented result due to using the charm cross-section calculation at LO only. The ERS dipole model prediction~\cite{Enberg:2008te}, that is mostly used by the experimental collaborations in their data analysis, is also consistent with the prediction of this paper, within uncertainties. The uncertainties in the ERS prediction are smaller compared to the QCD-based prediction, however the way of the uncertainty estimate in both calculations can not be directly compared. Indeed the dipole approach is expected to effectively resum logarithmic contributions of the form $\alpha_S\mathrm{ln}(1/x)$ in the PDF evolution, a property which could lead to a reduction of the PDF uncertainties associated to the target parton at low $x$, whereas the resummation of these logarithms is not included in the DGLAP evolution.
The uncertainties associated to the projectile parton distributions in Ref.~\cite{Enberg:2008te} were estimated by comparing two different central PDF sets, without considering the PDF uncertainty associated to each of those. Furthermore, the factorisation scale variation in Ref.~\cite{Enberg:2008te} is performed in a limited range of $\mu_F$= $m_c$, $\mu_F$=2 $m_c$.

\begin{figure}
\centering
    \includegraphics[width=0.49\textwidth]{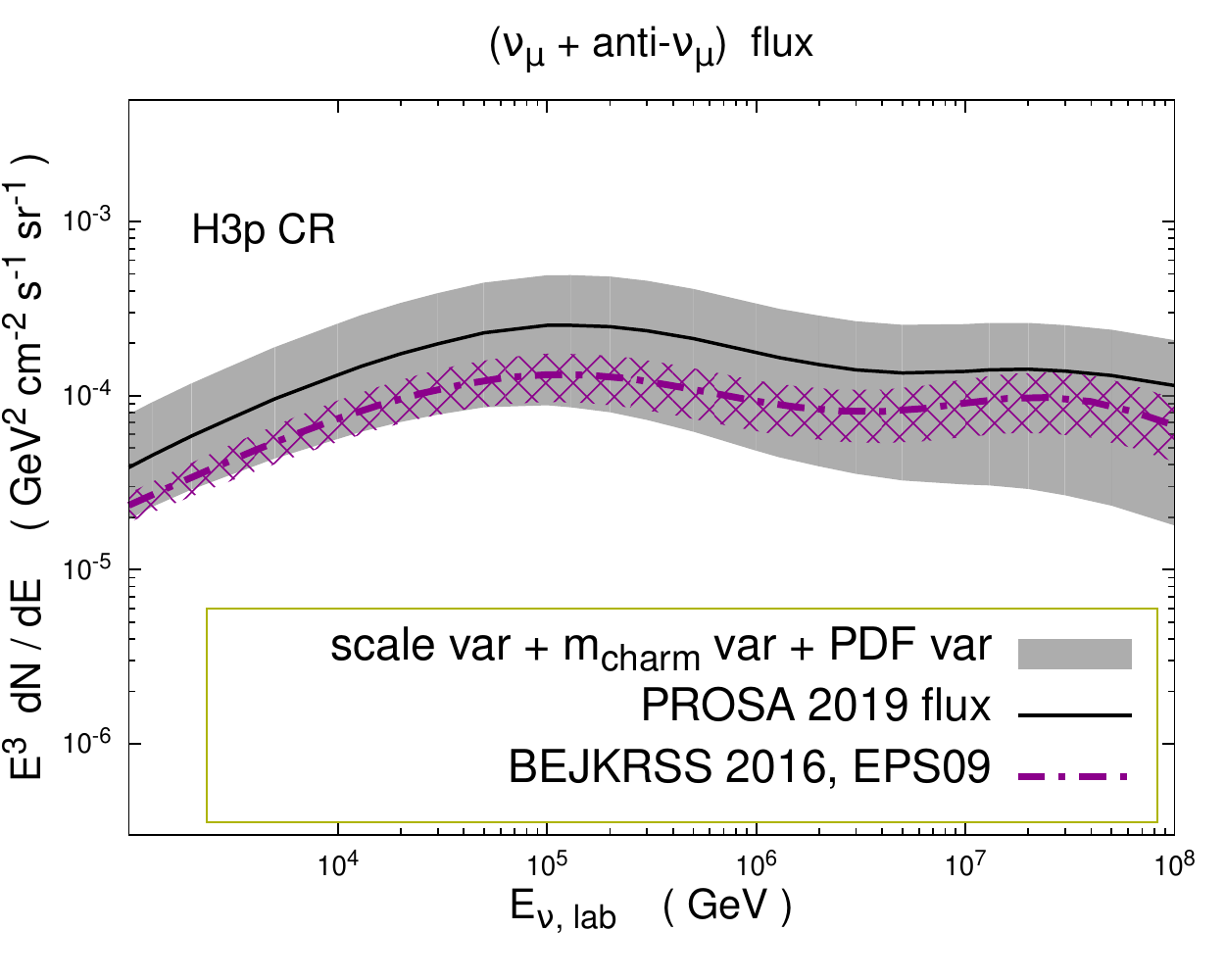}
    \includegraphics[width=0.49\textwidth]{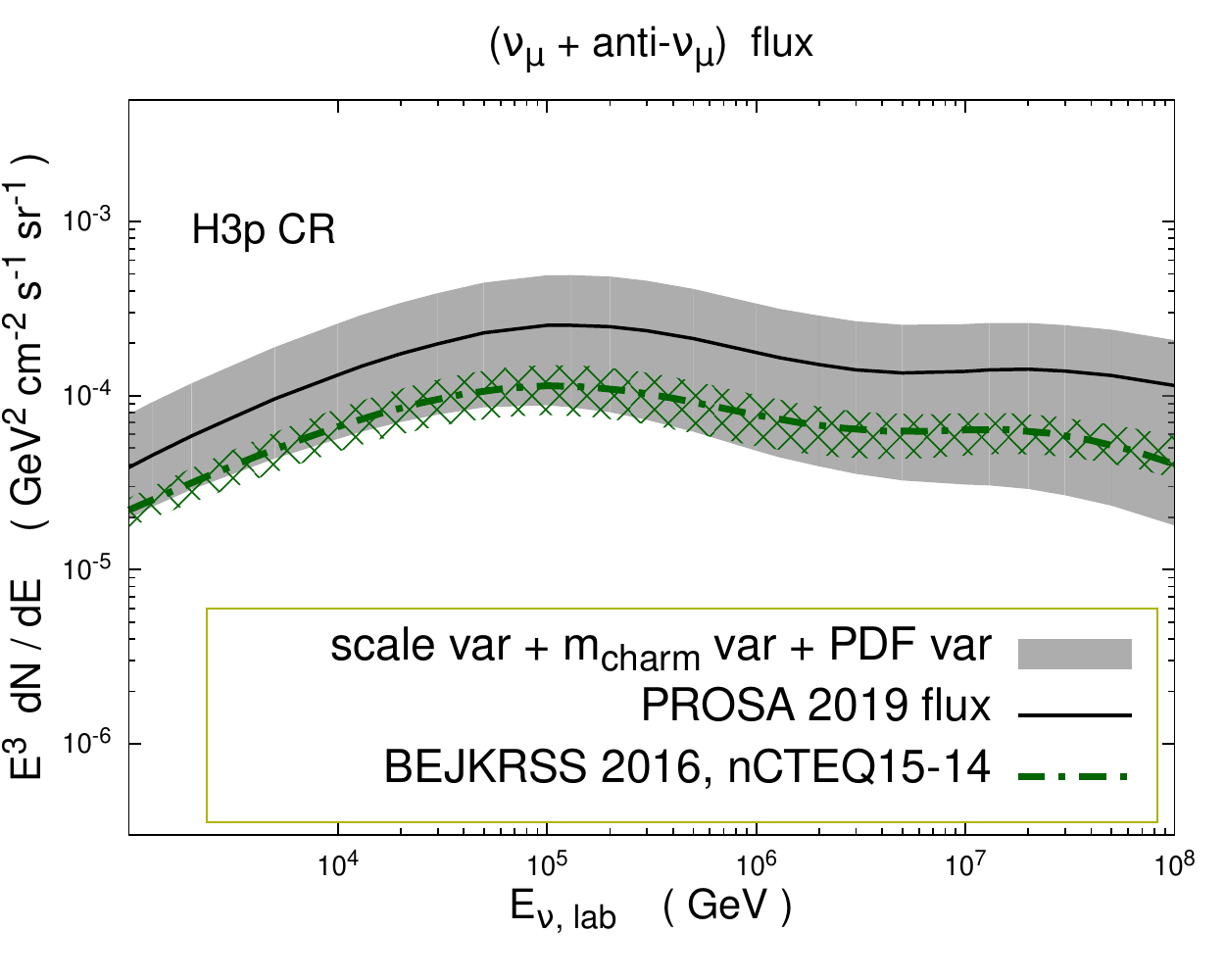}
  \caption{\label{fig5prompt} Predictions for prompt atmospheric-neutrino fluxes obtained in the presented analysis, compared to those by other authors~\cite{Bhattacharya:2016jce}. The all-nucleon flux H3p CR is used for consistency with Ref.~\cite{Bhattacharya:2016jce}. 
For the description of the Nitrogen (air) target, the nuclear PDFs EPS09~\cite{Eskola:2009uj} and nCTEQ15~\cite{Kovarik:2015cma} are used.}
\end{figure}

In Fig.~\ref{fig5prompt}, the prediction for the prompt neutrino flux based on the superposition model for both the projectile 
CR and the target nucleon of the air, obtained using the PROSA 2019 proton PDF set, is compared to the calculation of Ref.~\cite{Bhattacharya:2016jce} which uses nuclear PDFs to describe the target nucleon (nitrogen) and the proton PDFs for the projectile CR. The H3p CR all-nucleon spectrum is adopted as an input in our calculation, to be consistent with the choice of Ref.~\cite{Bhattacharya:2016jce}. In general, nuclear PDF fits are at a less advanced stage of development with respect to the proton PDF fits, due to the fact that less experimental data on collisions involving at least one nucleus are available with respect to the proton case, and that the theory for describing these collisions is also less advanced, with persistent difficulties in disentangling the different possible sources of cold nuclear matter effects. Additionally, the study of $p$-$A$ collisions at the Large Hadron Collider has been performed by mostly using Pb beams, while the atmosphere involves much lighter nuclei (N, O), which necessarily requires an important extrapolation.    
However, at present stage, it is remarkable to observe that predictions using proton PDFs and the superposition model turn out to agree with those using nuclear PDFs, at least within present uncertainty. 
This might point to the conclusion that 
the approximation of using proton PDFs and the superposition model, instead of nuclear PDFs (which, in principle, would be more appropriate because the air is made by nuclei instead of being made by unbound protons and neutrons) can still be considered as well justified, at least considering the present status of uncertainties. 

\section{Summary}
\label{sec:summary}

In this paper, improved constraints on the parton distributions are presented, as obtained in a QCD analysis at NLO using DIS and $pp$ collision data. In particular, the recent measurements of the LHCb and ALICE experiments of hadroproduction of charm and beauty-flavoured hadrons in different kinematic ranges (forward and central) provide additional sensitivity to the gluon distribution on a wide range of $x$ values. The assumptions on the initial parametrisation of the gluon distribution are investigated, which is important for understanding the low-$x$ behaviour of the gluon in the proton. For $x < 10^{-4}$, the gluon and sea quark PDFs turn out to be consistent and have smaller uncertainties with respect to our previous fit (PROSA 2015). The resulting PROSA2019 PDFs are extracted in FFNS and VFNS and can be used in e.g. high-energy astrophysical applications. In this paper, they  
 are used to obtain improved predictions for the prompt atmospheric-neutrino flux.
The neutrino flux predictions based on PROSA 2019 PDF set are consistent with the earlier PROSA results and have significantly improved accuracy. 
The presented neutrino flux predictions are also compatible with the calculations based on nuclear PDFs and with the results of the IceCube experiment. 
{With the prospect of new facilities such as the LHeC and EIC on the horizon, improved analyses of the nuclear PDFs are emerging, and these issues may warrant reexamination in the near future.}

\section*{Acknowledgements}

We would like to thank I.~Novikov and A.~Glazov for their help with developing and using new features of the xFitter framework, and V.~Bertone for his help with the APFEL library. We are grateful to M.~Benzke for having produced the predictions for $D$-meson hadroproduction used as a basis for the computation of prompt neutrino fluxes in the GM-VFNS framework. We thank S. Alekhin for useful discussions and comments. The work of O.~Z. has been supported by Bundesministerium f\"ur Bildung und Forschung (contract 05H18GUCC1). 

The authors are grateful to the Mainz Institute for Theoretical Physics (MITP)
of the DFG Cluster of Excellence PRISMA+ (Project ID 39083149), for its
hospitality and its partial support during the completion of this work.


\clearpage

\begin{thebibliography}{99}

\bibitem{Dokshitzer:1977sg}
Y.~L. Dokshitzer,
\newblock Sov. Phys. JETP {\bf 46}, 641 (1977).

\bibitem{Gribov:1972ri}
V.~N. Gribov and L.~N. Lipatov,
\newblock Sov. J. Nucl. Phys. {\bf 15}, 438 (1972).

\bibitem{Altarelli:1977zs}
G.~Altarelli and G.~Parisi,
\newblock Nucl. Phys. B {\bf 126}, 298 (1977).

\bibitem{Curci:1980uw}
G.~Curci, W.~Furmanski, and R.~Petronzio,
\newblock Nucl. Phys. B {\bf 175}, 27 (1980).

\bibitem{Furmanski:1980cm}
W.~Furmanski and R.~Petronzio,
\newblock Phys. Lett. B {\bf 97}, 437 (1980).

\bibitem{Moch:2004pa}
S.~Moch, J.~A.~M. Vermaseren, and A.~Vogt,
\newblock Nucl. Phys. B {\bf 688}, 101 (2004), arXiv:hep-ph/0403192.

\bibitem{Vogt:2004mw}
A.~Vogt, S.~Moch, and J.~A.~M. Vermaseren,
\newblock Nucl. Phys. B {\bf 691}, 129 (2004), arXiv:hep-ph/0404111.

\bibitem{Abramowicz:2015mha}
H1, ZEUS, H.~Abramowicz {\em et~al.},
\newblock Eur. Phys. J. {\bf C75}, 580 (2015), arXiv:1506.06042.

\bibitem{Zenaiev:2015rfa}
PROSA, O.~Zenaiev {\em et~al.},
\newblock Eur. Phys. J. {\bf C75}, 396 (2015), arXiv:1503.04581.

\bibitem{Gauld:2015yia}
R.~Gauld, J.~Rojo, L.~Rottoli, and J.~Talbert,
\newblock JHEP {\bf 11}, 009 (2015), arXiv:1506.08025.

\bibitem{Gauld:2016kpd}
R.~Gauld and J.~Rojo,
\newblock Phys. Rev. Lett. {\bf 118}, 072001 (2017), arXiv:1610.09373.

\bibitem{Bertone:2018dse}
V.~Bertone, R.~Gauld, and J.~Rojo,
\newblock JHEP {\bf 01}, 217 (2019), arXiv:1808.02034.

\bibitem{Aaij:2013mga}
LHCb, R.~Aaij {\em et~al.},
\newblock Nucl. Phys. {\bf B871}, 1 (2013), arXiv:1302.2864.

\bibitem{Aaij:2013noa}
LHCb, R.~Aaij {\em et~al.},
\newblock JHEP {\bf 08}, 117 (2013), arXiv:1306.3663.

\bibitem{Garzelli:2016xmx}
PROSA, M.~V. Garzelli {\em et~al.},
\newblock JHEP {\bf 05}, 004 (2017), arXiv:1611.03815.

\bibitem{Ball:2014uwa}
NNPDF, R.~D. Ball {\em et~al.},
\newblock JHEP {\bf 04}, 040 (2015), arXiv:1410.8849.

\bibitem{Cacciari:2015fta}
M.~Cacciari, M.~L. Mangano, and P.~Nason,
\newblock Eur. Phys. J. {\bf C75}, 610 (2015), arXiv:1507.06197.

\bibitem{Cacciari:1993mq}
M.~Cacciari and M.~Greco,
\newblock Nucl. Phys. {\bf B421}, 530 (1994), arXiv:hep-ph/9311260.

\bibitem{H1:2018flt}
H1, ZEUS, H.~Abramowicz {\em et~al.},
\newblock Eur. Phys. J. {\bf C78}, 473 (2018), arXiv:1804.01019.

\bibitem{Aaij:2015bpa}
LHCb, R.~Aaij {\em et~al.},
\newblock JHEP {\bf 03}, 159 (2016), arXiv:1510.01707,
\newblock [Erratum: JHEP05,074(2017)].

\bibitem{Aaij:2016jht}
LHCb, R.~Aaij {\em et~al.},
\newblock JHEP {\bf 06}, 147 (2017), arXiv:1610.02230.

\bibitem{Acharya:2017jgo}
ALICE, S.~Acharya {\em et~al.},
\newblock Eur. Phys. J. {\bf C77}, 550 (2017), arXiv:1702.00766.

\bibitem{Acharya:2019mgn}
ALICE, S.~Acharya {\em et~al.},
\newblock Eur. Phys. J. {\bf C79}, 388 (2019), arXiv:1901.07979.

\bibitem{Alekhin:2014irh}
S.~Alekhin {\em et~al.},
\newblock Eur. Phys. J. {\bf C75}, 304 (2015), arXiv:1410.4412,
\newblock see \url{http://xfitter.org}.

\bibitem{Botje:2010ay}
M.~Botje,
\newblock Comput. Phys. Commun. {\bf 182}, 490 (2011), arXiv:1005.1481.

\bibitem{Accardi:2016ndt}
A.~Accardi {\em et~al.},
\newblock Eur. Phys. J. {\bf C76}, 471 (2016), arXiv:1603.08906.

\bibitem{Alekhin:2017kpj}
S.~Alekhin, J.~Bl{\"u}mlein, S.~Moch, and R.~Placakyte,
\newblock Phys. Rev. {\bf D96}, 014011 (2017), arXiv:1701.05838.

\bibitem{Ball:2017otu}
R.~D. Ball {\em et~al.},
\newblock Eur. Phys. J. {\bf C78}, 321 (2018), arXiv:1710.05935.

\bibitem{Abdolmaleki:2018jln}
xFitter Developers' Team, H.~Abdolmaleki {\em et~al.},
\newblock Eur. Phys. J. {\bf C78}, 621 (2018), arXiv:1802.00064.

\bibitem{openqcdrad}
S.~Alekhin,
\newblock {``OPENQCDRAD''},
\newblock see \url{http://www-zeuthen.desy.de/~alekhin/OPENQCDRAD/}.

\bibitem{Mangano:1991jk}
M.~L. Mangano, P.~Nason, and G.~Ridolfi,
\newblock Nucl. Phys. B {\bf 373}, 295 (1992).

\bibitem{Dowling:2013baa}
M.~Dowling and S.-O. Moch,
\newblock Eur. Phys. J. {\bf C74}, 3167 (2014), arXiv:1305.6422.

\bibitem{Aaron:2008ac}
H1, F.~D. Aaron {\em et~al.},
\newblock Eur. Phys. J. {\bf C59}, 589 (2009), arXiv:0808.1003.

\bibitem{Chekanov:2008ur}
ZEUS, S.~Chekanov {\em et~al.},
\newblock JHEP {\bf 04}, 082 (2009), arXiv:0901.1210.

\bibitem{Nason:1999zj}
P.~Nason and C.~Oleari,
\newblock Nucl. Phys. {\bf B565}, 245 (2000), arXiv:hep-ph/9903541.

\bibitem{Alekhin:2012un}
S.~Alekhin, K.~Daum, K.~Lipka, and S.~Moch,
\newblock Phys. Lett. {\bf B718}, 550 (2012), arXiv:1209.0436.

\bibitem{Melnikov:2004bm}
K.~Melnikov and A.~Mitov,
\newblock Phys. Rev. {\bf D70}, 034027 (2004), arXiv:hep-ph/0404143.

\bibitem{Bonvini:2019wxf}
M.~Bonvini and F.~Giuli,
\newblock Eur. Phys. J. Plus {\bf 134}, 531 (2019), arXiv:1902.11125.

\bibitem{James:1975dr}
F.~James and M.~Roos,
\newblock Comput. Phys. Commun. {\bf 10}, 343 (1975).

\bibitem{Dulat:2015mca}
S.~Dulat {\em et~al.},
\newblock Phys. Rev. {\bf D93}, 033006 (2016), arXiv:1506.07443.

\bibitem{Harland-Lang:2014zoa}
L.~A. Harland-Lang, A.~D. Martin, P.~Motylinski, and R.~S. Thorne,
\newblock Eur. Phys. J. {\bf C75}, 204 (2015), arXiv:1412.3989.

\bibitem{Tanabashi:2018oca}
Particle Data Group, M.~Tanabashi {\em et~al.},
\newblock Phys. Rev. {\bf D98}, 030001 (2018).

\bibitem{Aaron:2009aa}
{H1 and ZEUS Collaborations},
\newblock JHEP {\bf 01}, 109 (2010), arXiv:0911.0884.

\bibitem{prosaweb}
{PROSA} web-site,
\newblock see \url{https://prosa.desy.de}.

\bibitem{Bertone:2013vaa}
V.~Bertone, S.~Carrazza, and J.~Rojo,
\newblock Comput. Phys. Commun. {\bf 185}, 1647 (2014), arXiv:1310.1394.

\bibitem{Forte:2010ta}
S.~Forte, E.~Laenen, P.~Nason, and J.~Rojo,
\newblock Nucl. Phys. {\bf B834}, 116 (2010), arXiv:1001.2312.

\bibitem{Bertone:2017ehk}
The xFitter Developers Team, V.~Bertone {\em et~al.},
\newblock Eur. Phys. J. {\bf C77}, 837 (2017), arXiv:1707.05343.

\bibitem{Chekanov:2002be}
ZEUS, S.~Chekanov {\em et~al.},
\newblock Phys. Lett. {\bf B547}, 164 (2002), arXiv:hep-ex/0208037.

\bibitem{Chekanov:2006xr}
ZEUS, S.~Chekanov {\em et~al.},
\newblock Nucl. Phys. {\bf B765}, 1 (2007), arXiv:hep-ex/0608048.

\bibitem{Abramowicz:2010cka}
ZEUS, H.~Abramowicz {\em et~al.},
\newblock Eur. Phys. J. {\bf C70}, 965 (2010), arXiv:1010.6167.

\bibitem{Aktas:2007aa}
H1, A.~Aktas {\em et~al.},
\newblock Phys. Lett. {\bf B653}, 134 (2007), arXiv:0706.3722.

\bibitem{Aaron:2010ac}
H1, F.~D. Aaron {\em et~al.},
\newblock Eur. Phys. J. {\bf C67}, 1 (2010), arXiv:0911.5678.

\bibitem{Chatrchyan:2012bja}
CMS, S.~Chatrchyan {\em et~al.},
\newblock Phys. Rev. {\bf D87}, 112002 (2013), arXiv:1212.6660,
\newblock [Erratum: Phys. Rev.D87,no.11,119902(2013)].

\bibitem{Sirunyan:2017azo}
CMS, A.~M. Sirunyan {\em et~al.},
\newblock Eur. Phys. J. {\bf C77}, 459 (2017), arXiv:1703.01630.

\bibitem{Sirunyan:2019zvx}
CMS, A.~M. Sirunyan {\em et~al.},
\newblock Submitted to: Eur. Phys. J.  (2019), arXiv:1904.05237.

\bibitem{Ball:2017nwa}
NNPDF, R.~D. Ball {\em et~al.},
\newblock Eur. Phys. J. {\bf C77}, 663 (2017), arXiv:1706.00428.

\bibitem{Gaisser:2016uoy}
T.~K. Gaisser, R.~Engel, and E.~Resconi,
\newblock {\em {Cosmic Rays and Particle Physics}} (Cambridge University Press,
  2016).

\bibitem{Collaboration:2011nsa}
ANTARES, M.~Ageron {\em et~al.},
\newblock Nucl. Instrum. Meth. {\bf A656}, 11 (2011), arXiv:1104.1607.

\bibitem{Gaisser:2014foa}
T.~Gaisser and F.~Halzen,
\newblock Ann. Rev. Nucl. Part. Sci. {\bf 64}, 101 (2014).

\bibitem{Adrian-Martinez:2016fdl}
KM3Net, S.~Adrian-Martinez {\em et~al.},
\newblock J. Phys. {\bf G43}, 084001 (2016), arXiv:1601.07459.

\bibitem{Mascaretti:2019uqn}
C.~Mascaretti and F.~Vissani,
\newblock JCAP {\bf 1908}, 004 (2019), arXiv:1904.11938.

\bibitem{Bhattacharya:2016jce}
A.~Bhattacharya {\em et~al.},
\newblock JHEP {\bf 11}, 167 (2016), arXiv:1607.00193.

\bibitem{Garzelli:2015psa}
M.~V. Garzelli, S.~Moch, and G.~Sigl,
\newblock JHEP {\bf 10}, 115 (2015), arXiv:1507.01570.

\bibitem{Gaisser:2011cc}
T.~K. Gaisser,
\newblock Astropart. Phys. {\bf 35}, 801 (2012), arXiv:1111.6675.

\bibitem{Gaisser:2013bla}
T.~K. Gaisser, T.~Stanev, and S.~Tilav,
\newblock Front. Phys.(Beijing) {\bf 8}, 748 (2013), arXiv:1303.3565.

\bibitem{Kachelriess:2019oqu}
M.~Kachelriess and D.~V. Semikoz,
\newblock (2019), arXiv:1904.08160.

\bibitem{Goncalves:2017lvq}
V.~P. Goncalves, R.~Maciula, R.~Pasechnik, and A.~Szczurek,
\newblock Phys. Rev. {\bf D96}, 094026 (2017), arXiv:1708.03775.

\bibitem{Nason:2004rx}
P.~Nason,
\newblock JHEP {\bf 11}, 040 (2004), arXiv:hep-ph/0409146.

\bibitem{Frixione:2007nw}
S.~Frixione, P.~Nason, and G.~Ridolfi,
\newblock JHEP {\bf 09}, 126 (2007), arXiv:0707.3088.

\bibitem{Sjostrand:2014zea}
T.~Sj{\"o}strand {\em et~al.},
\newblock Comput. Phys. Commun. {\bf 191}, 159 (2015), arXiv:1410.3012.

\bibitem{Benzke:2017yjn}
M.~Benzke {\em et~al.},
\newblock JHEP {\bf 12}, 021 (2017), arXiv:1705.10386.

\bibitem{Kneesch:2007ey}
T.~Kneesch, B.~A. Kniehl, G.~Kramer, and I.~Schienbein,
\newblock Nucl. Phys. {\bf B799}, 34 (2008), arXiv:0712.0481.

\bibitem{Aartsen:2016xlq}
IceCube, M.~G. Aartsen {\em et~al.},
\newblock Astrophys. J. {\bf 833}, 3 (2016), arXiv:1607.08006.

\bibitem{Gondolo:1995fq}
P.~Gondolo, G.~Ingelman, and M.~Thunman,
\newblock Astropart. Phys. {\bf 5}, 309 (1996), arXiv:hep-ph/9505417.

\bibitem{Enberg:2008te}
R.~Enberg, M.~H. Reno, and I.~Sarcevic,
\newblock Phys. Rev. {\bf D78}, 043005 (2008), arXiv:0806.0418.

\bibitem{Thoudam:2016syr}
S.~Thoudam {\em et~al.},
\newblock Astron. Astrophys. {\bf 595}, A33 (2016), arXiv:1605.03111.

\bibitem{Dembinski:2017zsh}
H.~P. Dembinski {\em et~al.},
\newblock PoS {\bf ICRC2017}, 533 (2018), arXiv:1711.11432,
\newblock [35,533(2017)].

\bibitem{Schroeder:2019agg}
F.~G. Schr{\"o}der,
\newblock PoS {\bf ICRC2019}, 030 (2019), arXiv:1910.03721.

\bibitem{Alekhin:2012ig}
S.~Alekhin, J.~Blumlein, and S.~Moch,
\newblock Phys. Rev. {\bf D86}, 054009 (2012), arXiv:1202.2281.

\bibitem{Bhattacharya:2015jpa}
A.~Bhattacharya, R.~Enberg, M.~H. Reno, I.~Sarcevic, and A.~Stasto,
\newblock JHEP {\bf 06}, 110 (2015), arXiv:1502.01076.

\bibitem{Gauld:2015kvh}
R.~Gauld, J.~Rojo, L.~Rottoli, S.~Sarkar, and J.~Talbert,
\newblock JHEP {\bf 02}, 130 (2016), arXiv:1511.06346.

\bibitem{Fedynitch:2015zma}
A.~Fedynitch, R.~Engel, T.~K. Gaisser, F.~Riehn, and T.~Stanev,
\newblock EPJ Web Conf. {\bf 99}, 08001 (2015), arXiv:1503.00544.

\bibitem{Eskola:2009uj}
K.~J. Eskola, H.~Paukkunen, and C.~A. Salgado,
\newblock JHEP {\bf 04}, 065 (2009), arXiv:0902.4154.

\bibitem{Kovarik:2015cma}
K.~Kovarik {\em et~al.},
\newblock Phys. Rev. {\bf D93}, 085037 (2016), arXiv:1509.00792.
\end{thebibliography}

\end{document}